\documentclass[1p]{elsarticle}

\usepackage{amsthm}
\usepackage{amssymb}
\usepackage{amsmath}
\usepackage{amsfonts}

\usepackage{float}
\usepackage{epsf}
\usepackage{hyperref}

\usepackage[scr=esstix,cal=boondox]{mathalfa} 

\usepackage{textcomp}

\usepackage{color, colortbl}
\definecolor{Gray}{gray}{0.9}

\usepackage{multirow}

\usepackage[ruled,vlined]{algorithm2e}

\theoremstyle{remark}	\newtheorem{theorem}{Theorem}
\theoremstyle{remark}	
\theoremstyle{remark}	

\usepackage[font=footnotesize]{subcaption}

\biboptions{sort&compress}

\begin{document}

\title{Back to Basics: Fast Denoising Iterative Algorithm}
% with Application to Speckle Suppression} %based on Convolutional Autoencoding} %to be changed

\author{Deborah Pereg$^1$}
\footnote{MIT MechE, Harvard School of Engineering and Applied Sciences}

%\author{Deborah Pereg$^{1,2,3}$, Martin Villiger $^{2,3}$\\
%$^{1}$MIT CSAIL, $^{2}$Wellman Center for Photomedicine MGH, $^{3}$Harvard Medical School\\
%\texttt{deborahp@csail.mit.edu, mvilliger@mgh.harvard.edu }

%\date{December 10, 2022}

%\maketitle

\begin{abstract} 
We introduce Back to Basics (BTB), a fast iterative algorithm for noise reduction.
Our method is computationally efficient, does not require training or ground truth data, and can be applied in the presence of independent noise, as well as correlated (coherent) noise, where the noise level is unknown. We examine three study cases: natural image denoising in the presence of additive white Gaussian noise, Poisson-distributed image denoising, and speckle suppression in optical coherence tomography (OCT). Experimental results demonstrate that the proposed approach can effectively improve image quality, in challenging noise settings. Theoretical guarantees are provided for convergence stability.
\vspace{5mm}

\end{abstract}

\maketitle

%\numberwithin{equation}{section}

\section{Introduction}

Image denoising, defined as removal of a zero-mean independent and identically distributed (i.i.d.) Gaussian noise from an image \cite{Elad:2023}, has been an extensively studied problem, achieving remarkable results. To a certain degree, some may claim that it is a solved problem. 
Nevertheless, many imaging applications (e.g., \cite{Torem:2023,Goodman:2007}) exhibit a non-linear noise model, in which the noise is correlated with the signal. In these cases, the additive i.i.d noise assumption collapses. 

Supervised learning methods demonstrate impressive results for image denoising (e.g., \cite{Chen:2016,Alsaiari:2019}). Many classic algorithms apply a minimum squared error (MMSE) or maximum a posteriori (MAP) estimator (e.g. \cite{Elad:2006,Weiss:2002,Elad:2023}). As denoising is inherently an ill-posed task (i.e., there can be more than one unique viable solution), MMSE estimators typically tend to average over these solutions, thus producing blurry results. Recent prior-based image restoration methods, such as regularization by denoising (RED) \cite{Romano:2017}, and plug and play (PnP) \cite{Bouman:2013}, require the knowledge of a fidelity term, in addition to a trained denoiser that fits a known level of noise. Namely, the objective is decoupled into the data ﬁdelity term and the
regularization terms, for use in a proximal optimization algorithm (e.g.,
ADMM). The denoiser replaces the proximal operator of the regularization term representing the MAP solution of
a denoising problem. 
Unsupervised methods, such as block-matching-3D (BM3D) \cite{Dabov:2006} and non-local-means (NLM) \cite{buades:2011} perform well under the assumption of i.i.d noise, and in some cases also in the presence of coherent noise, such as for speckle suppression in medical imaging \cite{Yu:2016,Cuartas:2018}. Unfortunately, many of these methods exhibit a relatively high computational complexity, that may significantly hinder any real time application incorporating a denoiser as a step prior to further image analysis or downstream tasks. 

\begin{figure*}[t]
    \begin{subfigure}[t]{0.249\textwidth}
        \includegraphics[width=\linewidth]{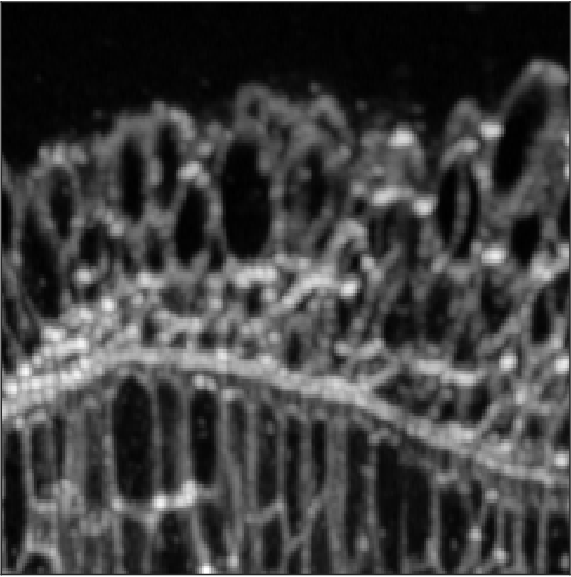}
				\caption{}
    \end{subfigure}%
		\hfill 
		\begin{subfigure}[t]{0.249\textwidth}
        \includegraphics[width=\linewidth]{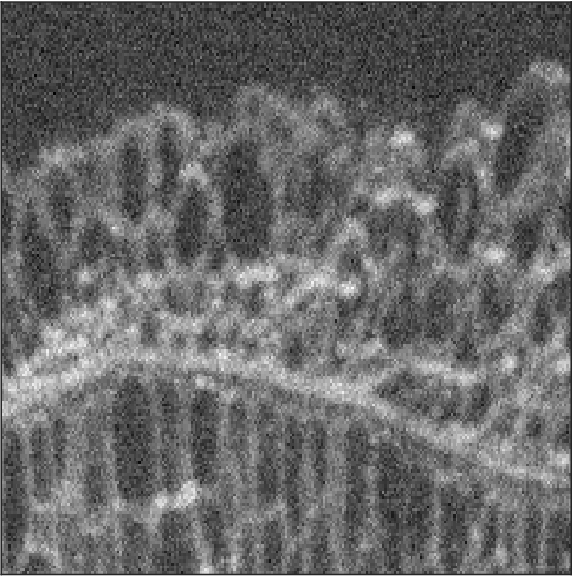}
				\caption{}
    \end{subfigure}%
		\hfill 
    \begin{subfigure}[t]{0.249\textwidth}
        \includegraphics[width=\linewidth]{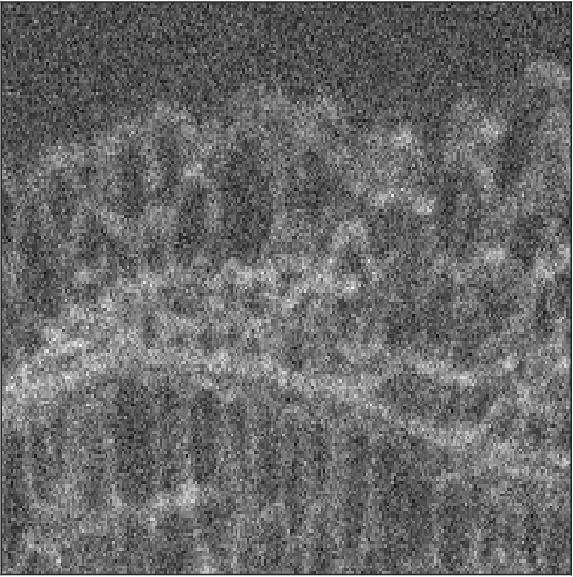}
				\caption{}
    \end{subfigure}%
		\hfill 
		\begin{subfigure}[t]{0.249\textwidth}
        \includegraphics[width=\linewidth]{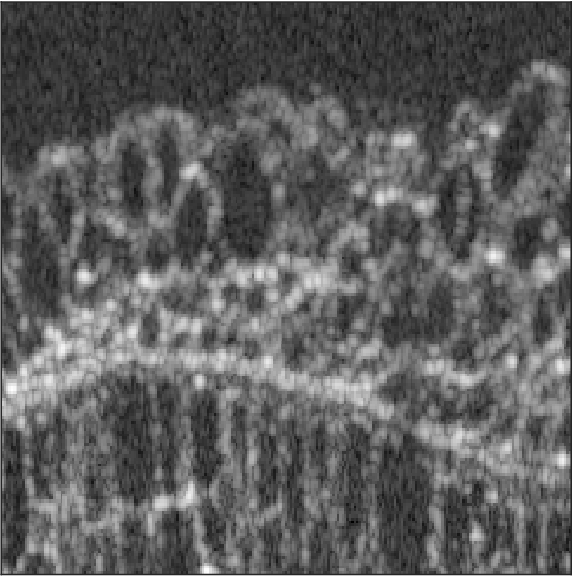}
				\caption{}
    \end{subfigure}%
\caption{Three types of noise. Visualization for cucumber OCT tomogram: (a) Cucumber OCT cross sectional ground truth example \cite{Pereg:2022}; (b) AWGN with variance $\sigma_{\mathrm{w}}=10$; (c) Poisson noise corrupted image; (d) Speckled tomogram.}
\label{fig_intro}
\end{figure*}

It is safe to say that additive white Gaussian noise (AWGN), Poisson noise and speckle noise are considered the three most critical types of real-world image noises (see Figure~\ref{fig_intro}). In digital images, sensor noise, thermal noise, electric-circuit noise, and quantization noise are often modeled as AWGN. Additionally, imaging noise arising from various physical causes is modeled as Gaussian noise due to the central limit theorem. As optical sensors are photon counting, Poisson noise is associated with the particle nature of light \cite{Costantini:2004}. Poisson noise is particularly dominant in optical imaging applications that require low light levels such as biomicroscopy. Lastly, speckle noise is critical in coherent imaging systems such as ultrasound and radar \cite{Goodman:2007}.

\subsection*{Related Work.}
Most denoising methods assume AWGN of a known noise level. Blind denoising, i.e., the removal of AWGN of unknown level has been investigated in several recent works.
Some of these works employ a neural network trained with examples of randomly varying noise levels, e.g., denoising convolutional neural networks (DnCNNs)\cite{Zhang:2017,Jain:2008}. Other works proposed to first estimate the noise level from the data and then employ a denoiser that is designed to fit the predicted noise level \cite{Liu:2013,Lebrun:2014}. An alternative approach, related to our work here, is to employ boosting of image denoising \cite{Romano:2015,Chen:2018}. That is, iteratively strengthen the signal by adding the previous denoised image to the degraded input image, denoise the strengthened image, and subtract the previous denoised image from the restored signal-strengthened outcome.

A recent line of work, somewhat related with RED and PnP, inspired by annealed Langevin dynamics, employs a trained denoiser as an estimator for the posterior distribution, thus iteratively sampling from the posterior distribution to obtain improved perceptual quality. This approach was recently employed for generative purposes (denoising diffusion probabilistic models \cite{Sohl:2015,Ho:2020,Song:2020}) as well as for score-based stochastic denoising \cite{Kawar:2021} and image reconstruction \cite{Simoncelli:2021,Milanfar:2023} in an attempt to avoid the so-called "regression to the mean'' phenomena.
Denoising diffusion reverses a known analytically defined degradation process, thus generating new samples starting from a fully degraded image (e.g., pure noise). 

As imaging sensors essentially count photons, their imaging noise is characterized by a Poisson distribution \cite{Costantini:2004} (also known as shot-noise). Poisson-modeled noise reduction has been extensively studied, primarily for astronomical, biomedical and photographic imaging \cite{starck:2007,Dey:2006,Figueiredo:2010}. 
That said, the associated optimization problems, starting from the classical Richardson-Lucy algorithm \cite{starck:2007} to more recent works involving inverse Poisson problems \cite{Torem:2023,Ron:2016}, are often computationally complex and slow.

Speckle is a type of interference, forming high-contrast patterns with grains-like appearance, that characterizes optical measurements of signal intensity, as well as ultrasound imaging, OCT, and radar. Speckle is not an additive statistically independent noise, but rather unresolved spatial information originating in the interference of many sub-resolution spaced scatterers \cite{Curatolo:2013}. Specifically, OCT tomograms display the intensity of the scattered light (as the log-valued squared norm of the complex-valued tomogram), where it is assumed that the contributions from structural features beyond the imaging resolution add up coherently to generate random speckle patterns \cite{Schmitt:1999,Goodman:2007}. Generally speaking, speckle appears in a signal when the signal is a linear combination of
independently random phased additive complex elements. 
The resulting sum is a random
walk, that may exhibit constructive or destructive interference depending on the
relative phases. The intensity of the
observed wave is the squared norm of this sum. Exact analogs of the speckle phenomenon appear in many other fields and applications. For example, the squared magnitude of the finite-time Fourier
transform (FFT) (the periodogram) of a sample function of almost any random process shows
fluctuations in the frequency domain that have the same single-point statistics
as speckle \cite{Porat:1996}.

For fully developed speckle, each pixel's value encapsulates a sum of a large enough number of reflectors, which,
according to the central limit theorem, has a Gaussian distribution. 
In this case, assuming a uniform phase, 
the intensity is distributed according to an exponential probability density,
\vspace{-5pt}
\begin{equation}
p_{Y|X}(y|x)=\frac{1}{x}\exp \Big(-\frac{y}{x}\Big), \quad y>0
\label{2.4}
\end{equation}
where $y$ is the measured intensity pixel value, and $x$ is the mean intensity, defining the ground truth. 
That is, $x=E_{Y|X}\big\{Y|X=x\big\}$, is the expected value over all speckle possible realizations for a given ground truth value. As the variance of (\ref{2.4}) is $x$, the fluctuations of fully developed speckle can be particularly destructive and visually disturbing, as they are of the order of a given ground truth pixel-value.  
Speckle that is not fully developed would have more complicated distributions' formulation, depending on the number of phasors, and their amplitudes and phases distribution. 

% Speckle suppression. 
Speckle suppression is often accomplished by incoherent averaging of images with different speckle realizations \cite{Pircher:2003}, e.g., through angular compounding \cite{Desjardins:2007, Zhao:2020}. Neverthelss, averaging methods tend to produce blurred images. Moreover, although effective at suppressing speckle in ex vivo tissues or in preclinical animal research, in practice obtaining multiple speckle realizations requires additional time and data throughput, which renders these approaches incompatible with clinical in vivo imaging.

Consequently, attempts to computationally suppress speckle has also been investigated. The majority of these algorithms - such as non-local means (NLM) \cite{Yu:2016,Cuartas:2018}, and BM3D \cite{chong:2013} - apply a denoiser under the (incorrect) assumption of i.i.d Gaussian noise. The solution is often sensitive to parameters' fine-tuning. Some algorithms also rely on accurately registered volumetric data, which is challenging to obtain in clinical settings. 

Recently, the use of supervised learning for speckle reduction has been extensively investigated \cite{shi:2019,devalla:2019,gour:2020,Dong:2020,ma:2018}. Unfortunately most supervised learning data-driven methods require a large training dataset, and are sensitive to the training source domain.  Furthermore, changes in the physical system's acquisition parameters may lead to degraded performance. To overcome this challenge, Pereg (2023) \cite{Pereg:2023B} proposed a domain-aware patch-based approach to train a model for speckle suppression with limited ground-truth data. Generally speaking, in medical imaging trained models are domain-dependent, and fail to generalize well under varying imaging systems, with different sampling rates, resolutions and contrasts. Thus, clearly, an unsupervised solution would be ideal. 

\subsection*{Contributions.}
Overall, the problem of denoising has been thoroughly investigated and there are abundant works in different fields and applications \cite{Elad:2023}. That said, \textit{most models assume that the degradation is known, as well the noise level.}
In this work we propose an iterative algorithm of relatively low computational complexity for noise reduction, under the assumption of an unknown noise level.
The proposed iteration update-rules are inherently defining a fixed-point successive iteration method and simple iteration method \cite{Banach:1922,Williamson:1973}, and as such, we provide theoretical analysis for their stable convergence under certain conditions. 
We demonstrate the applicability of our method for AWGN noise mitigation, Poisson noise, and speckle suppression in OCT. For speckle reduction, we employ an operation termed receptive-field-normalization (RFN) \cite{Pereg:2022} that reveals the zero-crossing patterns (optical vortices) visible in speckle patterns \cite{Goodman:2007}. 

The remainder of the paper is organized as follows. In Section \ref{sec2} we present background and problem formulation. Section \ref{sec3} describes the proposed iterative procedure and theoretical guarantees. 
Section \ref{sec4} presents the speckle-oriented denoiser based on RFN. Section \ref{sec5} presents experimental results. Finally we summarize and conclude our work in Section \ref{sec6}.

\section{Preliminaries}\label{sec2}
\subsection{Inverse Problems}
The inversion problem traditionally attempts to recover an unknown signal $\mathbf{x}\in\mathbb{R}^{n}$, given an observation $\mathbf{y}\in\mathbb{R}^{d}$ \cite{Elad:2023,Daubechies:2004}. The corresponding optimization problem is therefore formulated as
\begin{equation}\label{1.1}
\hat{\mathbf{x}}= \underset{\mathbf{x}\in\mathbb{R}^n}{\mathrm{argmin}}\, \ell(\mathbf{x},\mathbf{y})+\lambda \varphi(\mathbf{x}),
\end{equation}
where $\ell(\mathbf{x},\mathbf{y})$ is the fidelity term, $\varphi(\mathbf{x})$ is a regularization term, and $\lambda>0$ is a balancing weight. It is often assumed that $\mathbf{y}=\mathbf{Ax+w}$, where $\mathbf{A}\in\mathbb{R}^{d \times n}$ is a known linear degradation and $\mathbf{w}$ is an i.i.d white Gaussian noise (WGN). 
In that case the discrepancy is $\ell(\mathbf{x},\mathbf{y})=\frac{1}{2}\| \mathbf{y-Ax}\|_2^2$.
The linear model is employed in many tasks, such as: compressed senising, deconvolution, deblurring and super-resolution. The denoising linear model assumes $\mathbf{A}=\mathbf{I}$, where $\mathbf{I}$ is the identity matrix. 

\subsection{Problem Formulation}
We assume an input $\mathbf{y} \in \mathbb{R}^{n}$  that is a corrupted version of $\mathbf{x} \in \mathbb{R}^{n}$, such that
\begin{equation}\label{1.2}
\mathbf{y}=u(\mathbf{x})+v(\mathbf{x}),
\end{equation}
where $u(\cdot)$ and $v(\cdot)$ are \textit{unknown} functions.
Typically, $u(\cdot)$ describes the operation of a deterministic system on the source signal, and $v(\mathbf{x})$ is an additional random noise term, that could depend on $\mathbf{x}$.
Our task is to recover $\mathbf{x}$. That is, we're attempting to find an estimate $\hat{\mathbf{x} }$ of the unknown ground truth $\mathbf{x}$. 
Most existing works, however, assume that the noise is uncorrelated with the signal.
We will take a more pragmatic approach, aiming to provide a speed-up solution that is a good approximation of the ground truth. 
We reformulate (\ref{1.2}),
\begin{equation}\label{1.3}
\mathbf{y}=\mathbf{x}+\mathbf{w},
\end{equation}
where $\mathbf{w} \in \mathbb{R}^{n}$ is an additional ``noise" (error) term.
The noise energy level $E\|\mathbf{w}\|^2$ is \textit{unknown}. $E$ denotes mathematical expectation.
In (\ref{1.3}), $\mathbf{w}$ essentially ``swallows" any form of corruption of $\mathbf{x}$ described in (\ref{1.2}).
That is, $\mathbf{w}$ encapsulates any form of deviation from the clean signal $\mathbf{x}$, regardless of its origin. 
We do \textit{not} assume that $\mathbf{w}$ is neither i.i.d nor that it is uncorrelated with $\mathbf{x}$.

%\subsection{Iterative Thresholding Algorithms}
%
%\paragraph{Algorithm Unfolding}

\subsection{Regularization by denoising (RED) \cite{Romano:2017}}
Romano et al. \cite{Romano:2017} introduced a framework based on the optimization 
\begin{equation}\label{1.4}
\hat{\mathbf{x}}_{\mathrm{RED}} =  \underset{\mathbf{x}\in\mathbb{R}^n}{\mathrm{argmin}} \quad \ell(\mathbf{x},\mathbf{y})+\frac{\lambda}{2}\big<\mathbf{x}\ , \mathbf{x}-f(\mathbf{x})\big>,
\end{equation}
where $f(\cdot)$ is a denoiser. 
Generally speaking, the function $f: \mathbb{R}^d\rightarrow\mathbb{R}^n$ maps the input measurement to its corresponding denoised signal $\hat{\mathbf{x}}=f(\mathbf{y})$, ideally $\hat{\mathbf{x}}=\mathbf{x}$.
A small value obtained for the regularization term $\big<\mathbf{x}\ , \mathbf{x}-f(\mathbf{x})\big>$ indicates either that the denoising residual $\mathbf{x}-f(\mathbf{x})$ is very small, implying that the recovered image is approaching a fixed point solution, or that the cross-correlation of the image and the residual is very small, implying that the residual is orthogonal to the image manifold, and, as such, exhibits white noise behavior.
The authors proposed several iterative algorithms to solve (\ref{1.4}) - steepest descent, fixed-point iteration and ADMM - guaranteed to converge to the global minimum. 
Cohen et al. \cite{Cohen:2021} formulated the inverse problem
\begin{equation}\label{1.5}
\hat{\mathbf{x}}_{\mathrm{RED-PRO}} =  \underset{\mathbf{x}\in\mathbb{R}^n}{\mathrm{argmin}} \quad \ell(\mathbf{x},\mathbf{y})\, \quad s.t. \quad \mathbf{x} = f(\mathbf{x}).
\end{equation}
For a linear distortion $\mathbf{y}=\mathbf{Hx+w}$.
The solution of (\ref{1.4}) via steepest (gradient) descent (RED-SD) is formulated via the update rule
\begin{equation}\label{1.6}
\mathbf{x}_{t+1} =  \mathbf{x}_t -\frac{\mu_t}{1+\lambda}\Big(\nabla \ell(\mathbf{x}_t,\mathbf{y}) +\lambda \big(\mathbf{x}_t-f(\mathbf{x}_t)\big) \Big),
\end{equation}
where $\mu_t>0$ is the step size. Notably, at iteration $t>1$, the denoiser $f(\cdot)$ is \textit {no longer operating on a signal distorted by additive i.i.d WGN}. 
In practice, for the linear degradation model, the term $\nabla \ell(\mathbf{x}_t,\mathbf{y}) = \mathbf{H}^T(\mathbf{Hy}-\mathbf{x}_t)$ re-injects a colored noise (residual) into the updated prediction at every iteration $t$.
Thus a possible intuitive explanation for the remarkable success of this strategy is that the solution is achieved via weighted averaging of different $T$ noisy realizations, combined with the incorporation of powerful denoisers such as NLM and TNRD.

\section{Back to Basics: Fast Iterative Denoiser}
\label{sec3}

\subsection{Fast Iterative Denoising} %\section{Proposed Method}
\textit{Assumption 1.}
Our first fundamental assumption, differing this model from its predecessors, is that we do not have access to a forward linear or non-linear model. In other words, $u(\mathbf{x})$ (in (\ref{1.2})) is unknown.
Therefore, $\ell(\mathbf{x},\mathbf{y})$ is unknown as well. In that case, one may suggest, that given the reformulation in (\ref{1.3}), we can still use $(\ref{1.4})$ in a similar manner, such that $\mathbf{y}=\mathbf{x+w}$. Unfortunately, in practice, we observe that the term $\nabla \ell(\mathbf{x}_t,\mathbf{y}) = \mathbf{y}-\mathbf{x}_t$ is re-injecing the residual at every iteration, which causes instability, significantly under low SNRs and correlated noise, such as speckle noise. \\
\textit{Assumption 2.}
Assume a denoiser $f(\mathbf{x})$.
The optimal solution obeys $\mathbf{x}^* \in \mathrm{Fix}(f)$, where
\begin{equation}\label{3a}
\mathrm{Fix}(f) \triangleq \{\mathbf{x} \in \mathbb{R}^n \ : \  f(\mathbf{x})=\mathbf{x} \}.
\end{equation}
In other words, a denoiser operating on the clean image rests. In practice, this is hard to verify.
To circumvent this limitation, we can relax this assumption to $\mathbf{x}^* \in \mathrm{Fix}_\epsilon(f)$ \cite{Cohen:2021}, such that for $\epsilon>0$ the $\epsilon$-approximate fixed-points
\begin{equation}\label{3.0}
\mathrm{Fix}_\epsilon(f) \triangleq \{\mathbf{x} \in \mathbb{R}^n \ : \ \| \mathbf{x} - f(\mathbf{x})\|\leq\epsilon \}.
\end{equation}

Our proposition is therefore, quite simple and intuitive. 
At iteration $t$, we simply re-apply the denoiser to the previous output. Namely,
\begin{equation}\label{3.1}
\mathbf{x}_{t+1} =  f(\mathbf{x}_t).
\end{equation}
Alternatively, considering a hybrid Banach contracting principle \cite{Banach:1922,Yamada:1998}, we can formulate the update rule as,
\begin{equation}\label{3.2}
\mathbf{x}_{t+1} =  \mathbf{x}_t -\mu_t\big(\mathbf{x}_t-f(\mathbf{x}_t)\big)=
(1-\mu_t) \mathbf{x}_t +\mu_t f(\mathbf{x}_t),
\end{equation}
where $\mu_t \in (0,1]$ is a step size, as described in Algorithms \ref{alg1}.
Another possible option focuses on the observation (input) as an anchor point,
\begin{equation}\label{3.3}
\mathbf{x}_{t+1} =  \mathbf{x}_0 -\mu_t\big(\mathbf{x}_0-f(\mathbf{x}_t)\big)=
(1-\mu_t) \mathbf{x}_0 +\mu_t f(\mathbf{x}_t), \qquad \mathbf{x}_0=\mathbf{y}.
\end{equation}

It is possible to add a noise term $\mathbf{e}_t\sim\mathcal{N}(0,\mathbf{I})$, similarly to Langevin dynamics algorithm \cite{Roberts:1996,Besag:2001}, such that the update rule is
\begin{equation}\label{3.4}
\mathbf{x}_{t+1} =  (1-\mu_t) \mathbf{x}_t +\mu_t f(\mathbf{x}_t) + \beta \mathbf{e}_t , \qquad \mathbf{x}_0=\mathbf{y},
\end{equation}
where $\beta$ is an appropriately small constant. The added term $\mathbf{e}_t$ enables
stochastic sampling, avoiding mode collapse. However, in our study-cases we did not observe significant improvement because in practice very few iterations are required to reach a solution. 

\vspace{0.5cm}
\SetKwInput{kwInit}{Init}
\begin{algorithm}[H]
\SetCustomAlgoRuledWidth{0.45\textwidth}
\SetAlgoLined
\SetKwInOut{Input}{input}\SetKwInOut{Output}{output}
\Input{Input image $\mathbf{y}\in\mathbb{R}^{n}$, 
denoiser $f(\cdot)$, $\{\mu_t\}_{t \in \mathbb{N}} \in(0,1], T>0$}
\kwInit{$\mathbf{x}_0=\mathbf{y}$} 

	\While{
	$\|\mathbf{x}_{t+1}-\mathbf{x}_t\|_2< \delta $ or $t \leq T$}{

		$\bullet \quad  \mathbf{z}_{t+1}=f(\mathbf{x}_{t})$\\
		$\bullet \quad  \mathbf{x}_{t+1}= (1-\mu_t)\mathbf{x}_{t}+ \mu_t \mathbf{z}_{t+1}$\\
		$\bullet \quad  t \leftarrow t+1$
	}
\Output{$\mathbf{x}_{t+1}$.}
\caption{Fast Iterative Denoising Algorithm (BTB)} 
\label{alg1}
\end{algorithm}

\subsection{Theoretical Analysis}
\paragraph{Fixed-Point Theory} We consider a nonlinear mapping $T:\mathbb{R}^n\rightarrow\mathbb{R}^n$, 
we say $\mathbf{x}$ is a fixed-point of $T$ if and only if $T(\mathbf{x})=\mathbf{x}$ \cite{Banach:1922}. 
The update steps (\ref{3.1}),(\ref{3.2}) define a sequence of iterates of the successive iteration method and the simple iteration method, respectively \cite{Williamson:1973}.
Generally speaking, by the classical Picard-Banach-Caccioppoli principle, the sequence in (\ref{3.1}) converges to a unique fixed point if $T$ is a strict contractive mapping. Namely, for some $q \in [0,1)$,
\begin{equation}
\|T(\mathbf{x})  -  T(\mathbf{y}) \|  \leq  q \| \mathbf{x}  -  \mathbf{y} \|, \ \forall \mathbf{x},\mathbf{y} \in \mathbb{R}^n. 
\end{equation}
While general denoisers cannot be guaranteed to form a strict contractive mapping, we can still show convergence of the above update step under certain conditions. 

%\begin{theorem}
%\label{Theorem 1} 
%Let $f(\mathbf{x})$ be a demicontructive denoiser \\
%Assume \\
%Then the sequence $\{\mathbf{x}_t\}_{t=0}^{T}$ generated by Algorithm 1, initialized with $\mathbf{x}_0=\mathbf{y}$, converges to a fixed point solution of the denoiser,
%\begin{equation}
%\hat{\mathbf{x}}=\{ \mathbf{x} \in \mathbb{R}^n : \mathbf{x}=f(\mathbf{x}) \},
%\end{equation}
%such that $\ell(\hat{\mathbf{x}},\mathbf{y}) \triangleq \frac{1}{2}\| \mathbf{y-x}\|_2^2$ obeys,
%\begin{equation}
%\ell(\hat{\mathbf{x}},\mathbf{y}) \leq \Delta.
%\end{equation}
%\end{theorem}

\begin{theorem}
\label{Theorem1} (Successive Iteration Method)
Let $\mathbf{x}_0 = \mathbf{y}$ be a noisy image such that, $\mathbf{y}=\mathbf{x}^*+\mathbf{w}$, where $\mathbf{x}^*$ is the noiseless image.
Define a sequence $\{\mathbf{x}_t\}_{t\leq T}$ by setting $\mathbf{x}_t=f(\mathbf{x}_{t-1})$,
where $f: \mathbb{R}^n \rightarrow \mathbb{R}^n$ is an ideal denoiser, such that
$\|\mathbf{w}_{t+1}\|\leq q \|\mathbf{w}_{t}\| \ , q \in[0,1)$, where $\mathbf{w}_t:=\mathbf{x}_{t}-\mathbf{x}^*$ . 
Then, the sequence $\{\mathbf{x}_t\}_{t=0}^{T}$, initialized with $\mathbf{x}_0=\mathbf{y}$, converges to a unique fixed point solution of the denoiser. The solution $\hat{\mathbf{x}}$ obeys
$\| \hat{\mathbf{x}}-\mathbf{y} \| \leq \| \mathbf{w} \|$.
\end{theorem}
 
\begin{theorem}
\label{Theorem2} (Simple Iteration Method)
Let $\mathbf{x}_0 = \mathbf{y}$ be a noisy image such that, $\mathbf{y}=\mathbf{x}^*+\mathbf{w}_0$, where $\mathbf{x}^*$ ideally is the noiseless image.
Define a sequence $\{\mathbf{x}_t\}_{t\leq T}$ by setting $\mathbf{x}_t=(1-\mu_t)\mathbf{x}_{t-1} + \mu_t f(\mathbf{x}_{t-1})$, $\mu_t \in (0,1)$,
where $f: \mathbb{R}^n \rightarrow \mathbb{R}^n$ is an ideal denoiser, such that 
$\|\mathbf{w}^\mathrm{f}_t\|\leq q \|\mathbf{w}^\mathrm{x}_t\| \ , q \in[0,1)$, where
$\mathbf{w}^\mathrm{x}_t:=\mathbf{x}_{t}-\mathbf{x}^*$ and $\mathbf{w}^\mathrm{f}_t:=f(\mathbf{x}_{t})-\mathbf{x}^*$.
Then, the sequence $\{\mathbf{x}_t\}_{t=0}^{T}$, initialized with $\mathbf{x}_0=\mathbf{y}$, converges to a unique fixed point solution of the denoiser. The solution $\hat{\mathbf{x}}$ obeys
$\| \hat{\mathbf{x}}-\mathbf{y} \| \leq \| \mathbf{w} \|$
\end{theorem}
\textit{Proofs.} See \ref{appA}. 

\begin{figure}[t]
\centering
\includegraphics[scale=0.65]{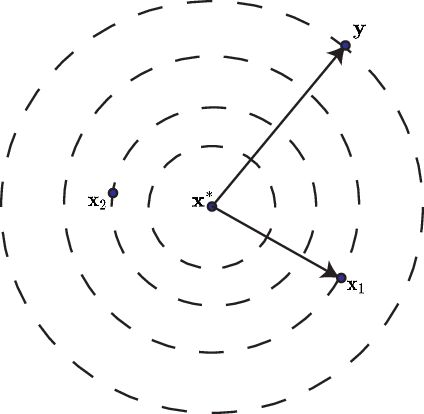} 
\caption{Illustration of BTB convergence.}
\label{fig0}
\end{figure}

In other words, the solution does not ``run too far"' from the measurement. Figure~\ref{fig0} illustrates the convergence process.
Note that we are not minimizing $\ell(\hat{\mathbf{x}},\mathbf{y})$, because we cannot assume the optimal solution is the one closest to the measurement. 
In practice, in some case, it may be challenging to verify that the condition $\|\mathbf{w}_{t+1}\|< q \|\mathbf{w}_{t}\| \ , q \in[0,1)$ holds. That is, that at every iteration the denoiser keeps reducing the noise (error) term energy, whether it is correlated with the signal or not. 
Nevertheless, as can be seen in the experiments below (Section \ref{sec5}), in practice, the proposed method is stable, and converges in very few iteration. 

\paragraph{Related work} Cohen et al. \cite{Cohen:2021} reformulated RED as a convex minimization problem regularized using the fixed-point set of demicontractive denoiser $f(\cdot)$. 
\begin{equation}\label{red-pro}
\hat{\mathbf{x}}_{\mathrm{RED-PRO}} =  \underset{\mathbf{x}\in\mathbb{R}^n}{\mathrm{argmin}} \quad \ell(\mathbf{x},\mathbf{y})\, \quad s.t. \quad \mathbf{x} \in \mathrm{Fix}(f).
\end{equation}
Under (\ref{red-pro}), it suffices to assume the denoiser is a demicontructive mapping. Although, as stated by the authors, demoicontractivity is difficult to prove for general denoisers, as other conditions.

Our work here shares with Delbracio et al. (2023) a distinct core observation underlying the proposed approach \cite{Milanfar:2023}: A small restoration step avoids the regression-to-the-mean effect because the set of
plausible slightly-less-bad images is relatively small. That said, we refrain from making this statement, as we believe it may be controversial. As although smaller steps lead to a smaller set of possible reconstructions, these slightly less noisy reconstructions are obtained by a denoising engine that often tends to yield slightly blurred results. Furthermore, when employing the simple iterative method, clearly we are averaging solutions, thus, we may still converge to an average of plausible reconstructions. To some extent, we may not be able to avoid this, as illustrated in Fig.~\ref{fig0}, the ideal solution is inherently modeled as the center of the noisy signals' spheres, and thus by definition should be the expected mean of our distorted observations. 

\section{Receptive Field Normalization (RFN) denoiser} \label{sec4}

We now turn to describe a specific denoiser that will be shown to be quite effective for speckle reduction.
Indeed, the RFN denoiser can be utilized for AWGN denoising as well, yet it is not the most effective one, in comparison with other known state-of-the-art denoisers.
Receptive field normalization (RFN) operator was first introduced in \cite{Pereg:2021} for a fast sparse coding (SC) algorithm for seismic reflectivity estimation. The fast SC algorithm is inspired by the classic iterative thresholding algorithms \cite{Daubechies:2004,Beck:2009}, and it produces a relatively good approximation of a convolutional sparse code (SC) under certain conditions. 
Additional details about RFN are in \ref{appB}. 

Let us briefly revise the definition of the one-dimensional (1D) RFN operation.\\
\textit{Definition 3: Receptive Field Normalization Kernel.}\\
A kernel $h[k]$ can be referred to as a receptive field normalization kernel if 
\begin{enumerate}
\item The kernel is positive: $h[k] \geq 0 \quad \forall k$.
\item The kernel is symmetric: $h[k]=h[-k] \quad \forall k$.
\item The kernel maximum is at its center: $h[0] \geq h[k] \quad \forall k \neq 0$.
\item The kernel's energy is finite $ \sum_k {h[k]} < \infty$.
\end{enumerate}
\textit{Definition 4: Receptive Field Normalization.}\\
Assuming a receptive field normalization kernel $h[k]$ of odd length $L_\mathrm{h}$, we define the local weighted energy of a time window centered around the $k$'th sample of a 1D observed data signal $\mathbf{v}\in\mathbb{R}^{L_\mathrm{v} \times 1}$
\begin{equation}\label{4.1}
\sigma_{\mathrm{v}}[k] \triangleq \Bigg( \sum_{n=-\frac{L_\mathrm{h}-1}{2}}^{\frac{L_\mathrm{h}-1}{2}} { h[n] v^2[k-n] } \Bigg) ^ \frac{1}{2} = \sqrt{h[k]*v^2[k]} .
\end{equation} 
Where $h[k]$ is a receptive field normalization window, and $*$ denotes the convolution operation. For our application we used a truncated Gaussian-shaped window, but it is possible to use any other window function depending on the application, such as: a rectangular window, Epanechnikov window, etc. The choice of the normalization window and its length affects the choice of the thresholding parameter. If $h[k]$ is a rectangular window, then $\sigma_\mathrm{v}[k]$ is simply the $\ell_2$ norm of a data segment centered around the $k$'th location. Otherwise, if the chosen receptive field normalization window is attenuating, then the energy is focused on the center of the receptive field, and possible events at the margins are repressed. 

Receptive field normalization is employed by dividing each point in the center of a local receptive field by its variance. Namely, we compute the local variance of $v[k]$ as defined in (\ref{4.1}). Then, we re-scale each point-value by dividing it with the energy of its receptive field, namely,
\begin{equation}\label{4.2}
\tilde{v}[k]=v[k]/\tilde{\sigma}_\mathrm{v}[k], 
\end{equation}
where $\tilde{\sigma}_\mathrm{v}[k]$ is a clipped version of $\sigma_\mathrm{v}[k]$ used in order to avoid amplification of low energy regions. Namely,
\begin{equation}\label{4.3}
\tilde{\sigma}_\mathrm{v}[k]=
\begin{cases}
\sigma_\mathrm{v}[k] &  \sigma_\mathrm{v}[k]  \geq \tau \\
1 &  \sigma_\mathrm{v}[k]  < \tau
\end{cases},
\end{equation}
where $\tau>0$ is a predetermined threshold. The extension to 2D is in \ref{appB}. 

\begin{figure}[t]
\centering
\includegraphics[scale=0.45]{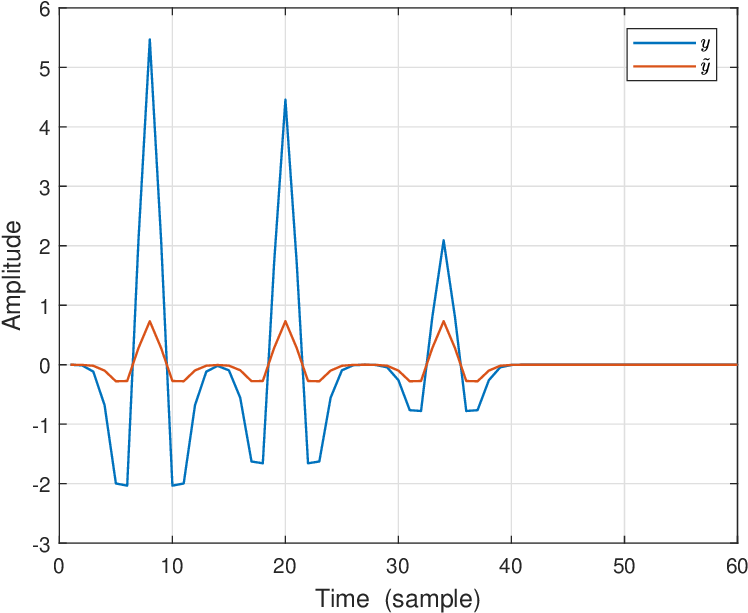} 
\caption{Synthetic data receptive field normalization example \cite{Pereg:2022}: (a) seismic trace $y$; (b) seismic trace after applying receptive-field normalization $\tilde{y}$, with an averaging window of size $L_\mathrm{h}=15$.}
\label{fig1}
\end{figure}

The term receptive field is borrowed from the study of the visual brain cortex, which inspired the term receptive field in CNNs, to describe a small limited local visual area that a neuron reacts to \cite{Geron:2017}. In CNNs, a neuron located in a certain layer is connected only to the output of neurons in a limited small area of the previous layer. Considering (\ref{4.1})-(\ref{4.2}), the normalization is with respect to a small local restricted support of $L_{\mathrm{h}}$ samples around each point in the data. We can refer to each of these small local support areas as a receptive field.

Figure~\ref{fig1} presents a simple synthetic example before and after receptive-field normalization, borrowed from \cite{Pereg:2022}. In this example, the pulses (Mexican-hat shaped) are intentionally completely separated. As can be clearly seen, when the pulses are sufficiently separated, the normalized signal is perfectly balanced regardless of the original local energy. Also, the pulses' shape is preserved. Each data point is scaled by its local neighborhood energy, but we still approximately keep the signal's shape.

\textit{Definition 5: RFN Operator.}\\ Let us define the following RFN operator $g(\cdot)$. Given an input signal $\mathbf{v}\geq 0$ as defined in (\ref{4.2}). Then,
\begin{equation}\label{4.4}
g(\mathbf{v}) = (\tilde{\mathbf{v}}-1)\odot \mathbf{v}
\end{equation} 
where $\odot$ denotes the Hadamard product.\footnote{for $\mathbf{v}\in \mathbb{R}^n$, (\ref{4.4}) can be replaced with $g(\mathbf{v}) = [\tilde{\mathbf{v}}-\mathrm{sign}(\mathbf{v})]\odot \mathbf{v}$, where $\mathrm{sign}(\cdot)$ denotes the signum function.}
%, and $\mathrm{sign}(\cdot)$
%is an element-wise operation such that,
%\begin{equation*}
%\mathrm{sign}(v)=
%\begin{cases}
%1  &  \quad if \  v \geq 0 \\
%-1 &  \quad if \  v > 0.
%\end{cases}
%\end{equation*}

\begin{figure*}[t]
    \begin{subfigure}[t]{0.24\textwidth}
        \includegraphics[width=\linewidth]{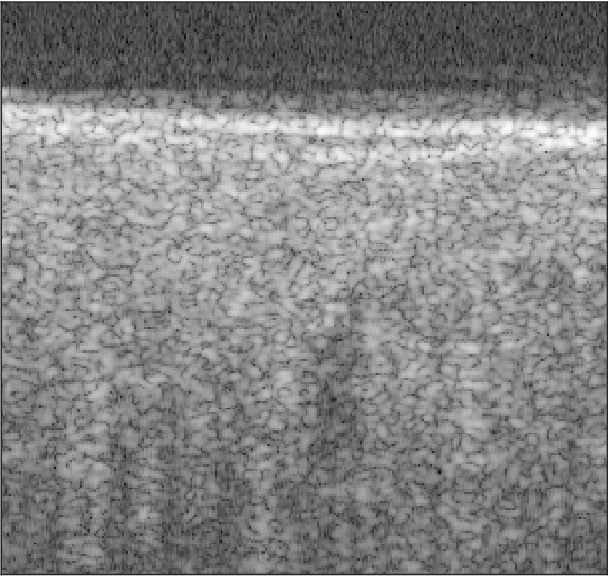}
				\caption{}
    \end{subfigure}%
		\hfill 
		\begin{subfigure}[t]{0.24\textwidth}
        \includegraphics[width=\linewidth]{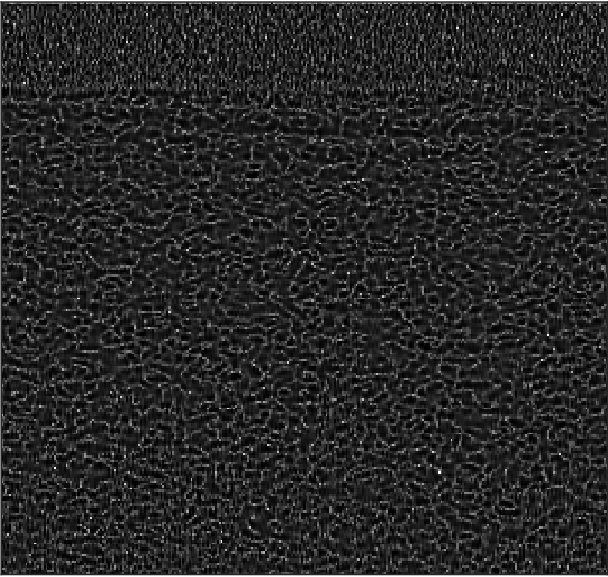}
				\caption{}
    \end{subfigure}%
		\hfill 
    \begin{subfigure}[t]{0.24\textwidth}
        \includegraphics[width=\linewidth]{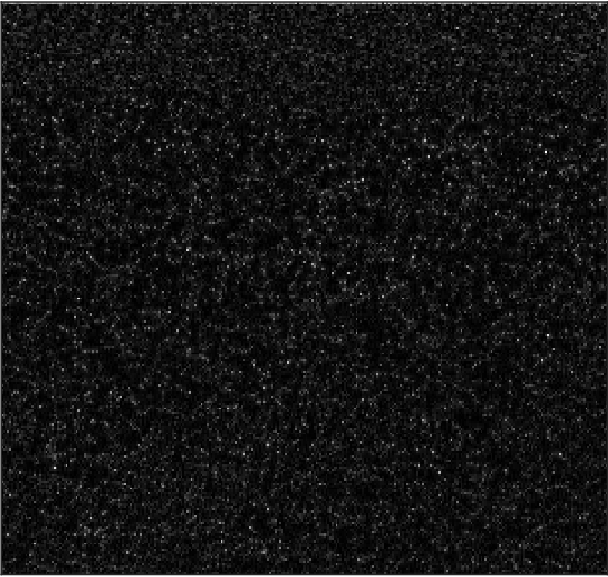}
				\caption{}
    \end{subfigure}%
		\hfill 
		\begin{subfigure}[t]{0.24\textwidth}
        \includegraphics[width=\linewidth]{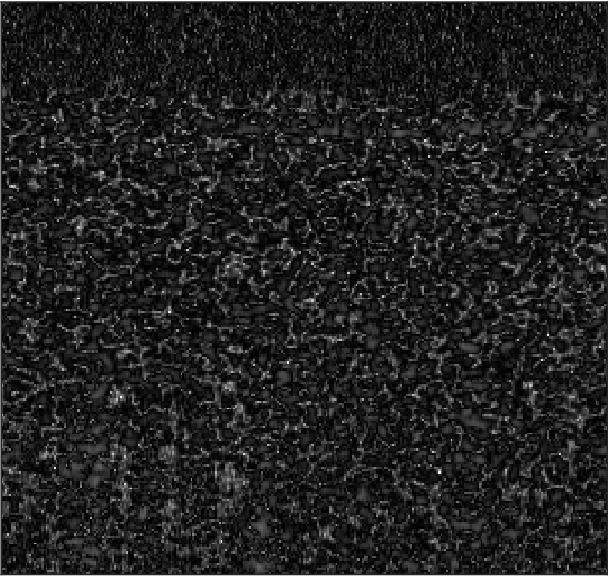}
				\caption{}
    \end{subfigure}%
\caption{RFN for OCT speckle visualization: (a) Chicken muscle OCT cross sectional example, $\mathbf{Y}$; (b) tomogram RFN result $|\tilde{\mathbf{Y}}-1|$ ; (c) High-pass filtered image $|\mathbf{Y}*\mathbf{H}_{\mathrm{Laplace}}|$ ; (d) $|\mathbf{Y}-\mathbf{X}|$ noise term.}
\label{fig2}
\end{figure*}

\subsection*{Zeros of Speckle Patterns: Optical Vortices}
The occurrence of zero intensity at a point in a speckle pattern
is an event that represents a more general phenomenon known as an optical singularity (or a wavefront
dislocation, or an optical phase vortex). Considerable amount of literature
was dedicated to the properties of such singularities (see, e.g., \cite{Nye:1974,Goodman:2007}).
Figure~\ref{fig2}(a) presents an example of a log-scaled speckled OCT cross-sectional tomogram of chicken muscle $\mathbf{Y}$.
As can be seen the zero crossings patterns are visually observable and to some degree are ``creating" the speckle grains interference. Figure~\ref{fig2}(b) presents the absolute value of $\mathbf{Y}$ after RFN operation $|\tilde{\mathbf{Y}}-1|$, compared with 
high-pass filtered image (Fig.~\ref{fig2}(c)) and with the estimated noise $\mathbf{Y}-\mathbf{X}$ (Fig.~\ref{fig2}(d)), where $\mathbf{X}$ is the clean ground truth. As can be visually observed the RFN operation efficiently decouples the zero-crossings and can, to some degree, serve as a good estimator of the noise term. 

Therefore, given the RFN operator $g(\cdot)$,
we can now define the RFN iterative denoising algorithm (as summarized in Algorithm \ref{alg2}), such that at iteration $t$, we simply reduce the detected zero-crossings speckle patterns extracted by the RFN operator applied on previous estimation. Namely, starting with $\mathbf{x}_0=\mathbf{y}$, 
\begin{equation}\label{4.5}
\mathbf{x}_{t+1} = \mathbf{x}_{t}- \alpha g(\mathbf{x}_{t}), \qquad \alpha \in (0,1],
\end{equation}
Considering a hybrid Banach contracting principle \cite{Banach:1922,Yamada:1998}, we can reformulate the update rule as,
\begin{equation}\label{4.6}
\mathbf{x}_{t+1} = (1-\mu) \mathbf{x}_t +\mu\big(\mathbf{x}_{t}- \tilde{\alpha} g(\mathbf{x}_{t})\big)=\mathbf{x}_{t}- \alpha g(\mathbf{x}_{t}), \qquad \alpha \in (0,1],
\end{equation}
where $\{\mu,\tilde{\alpha}\}\in(0,1]$ leading to an identical update rule as in (\ref{4.5}).
Hereafter we refer to this algorithm as Vortice. 

As the speckle zero-crossings may attenuate too fast at every iteration, a possible alternative of the update step above, which somewhat imitates the averaging of different speckle realizations, would reiterate over the first initial observation $\mathbf{y}$, as summarized in Algorithm \ref{alg4}.

\SetKwInput{kwInit}{Init}
\begin{algorithm}[H]
\SetCustomAlgoRuledWidth{0.45\textwidth}
\SetAlgoLined
\SetKwInOut{Input}{input}\SetKwInOut{Output}{output}
\Input{Input image $\mathbf{y}\in\mathbb{R}^{n}$, 
RFN-operator $g(\cdot)$, $\alpha \in(0,1],  \ T>0$}
\kwInit{$\mathbf{x}_0=\mathbf{y}$} 

	\While{
	$\|\mathbf{x}_{t+1}-\mathbf{x}_t\|_2< \delta $ or $t \leq T$}{

		$\bullet \quad  \mathbf{z}_{t+1}=g(\mathbf{x}_{t})$\\
		$\bullet \quad  \mathbf{x}_{t+1}= \mathbf{x}_{t}- \alpha \mathbf{z}_{t+1}$\\
		$\bullet \quad  t \leftarrow t+1$
	}
\Output{$\mathbf{x}_{t+1}$.}
\caption{RFN Iterative Denoising Algorithm (Vortice)} 
\label{alg2}
\end{algorithm}

%\SetKwInput{kwInit}{Init}
%\begin{algorithm}[H]
%\SetCustomAlgoRuledWidth{0.45\textwidth}
%\SetAlgoLined
%\SetKwInOut{Input}{input}\SetKwInOut{Output}{output}
%\Input{Input image $\mathbf{y}\in\mathbb{R}^{n}$, 
%RFN-operator $g_\beta(\cdot)$, $\alpha \in(0,1], \ T>0$}
%\kwInit{$\mathbf{x}_0=\mathbf{y}$} 
%
	%\While{
	%$\|\mathbf{x}_{t+1}-\mathbf{x}_t\|_2< \delta $ or $t \leq T$}{
%
		%$\bullet \quad  \mathbf{z}_{t+1}= g_\beta(\mathbf{x}_{t})$\\
		%$\bullet \quad  \mathbf{v}_{t+1}= \mathbf{x}_{t}- \alpha \mathbf{z}_{t+1}$\\
		%$\bullet \quad  \mathbf{x}_{t+1}= (1-\alpha) \mathbf{x}_{t}- \alpha \mathbf{v}_{t+1}$\\
		%$\bullet \quad  t \leftarrow t+1$
	%}
%\Output{$\mathbf{x}_{t+1}$.}
%\caption{Relaxed RFN Iterative Denoising Algorithm} 
%\label{alg3}
%\end{algorithm}
% Hybrid gradient descent method (HSD) - incorrect term

\SetKwInput{kwInit}{Init}
\begin{algorithm}[H]
\SetCustomAlgoRuledWidth{0.45\textwidth}
\SetAlgoLined
\SetKwInOut{Input}{input}\SetKwInOut{Output}{output}
\Input{Input image $\mathbf{y}\in\mathbb{R}^{n}$, 
RFN-operator $g(\cdot)$, $\{\alpha,\beta\} \in(0,1], \ T>0$}
\kwInit{$\mathbf{x}_0=\mathbf{y}$, $\mathbf{z}_0=\mathbf{0}$} 

	\While{
	$\|\mathbf{x}_{t+1}-\mathbf{x}_t\|_2< \delta $ or $t \leq T$}{

		$\bullet \quad  \mathbf{z}_{t+1}= g(\mathbf{x}_{t}) + \mathbf{z}_{t}$ \\
		$\bullet \quad  \mathbf{v}_{t+1}= \mathbf{y}-\beta \mathbf{z}_{t+1}$ \\
		$\bullet \quad  \mathbf{x}_{t+1}= (1-\alpha) \mathbf{x}_{t}+ \alpha \mathbf{v}_{t+1}$\\
		$\bullet \quad  t \leftarrow t+1$
	}
\Output{$\mathbf{x}_{t+1}$.}
\caption{RFN Iterative Speckle-Focused Denoising Algorithm} 
\label{alg4}
\end{algorithm}

%\footnote{It is possible to use $\{\mu_t\}_{t \in T}$ instead of $\alpha$.}. 

\subsection{Theoretical Analysis}
To provide some intuitive mathematical understanding of the RFN operation, we propose the following theorem.
\begin{theorem}
\label{Theorem4} 
Assume RFN operator $g(\mathbf{v})$ as defined in (\ref{4.4}), with $h$ rectangular window.
\begin{enumerate} 
\item If $\mathbf{v}=c\mathbf{1}$, where $\mathbf{1}$ is a vector of ones, $c>\tau$ is a constant, then $g(\mathbf{v})=\mathbf{0}$. 
\item If $\mathbf{v}[k]\sim \mathcal{N}(m_\mathrm{v}, s^2_\mathrm{v})$, then $E \sigma^2_\mathrm{v}[k] = s^2_\mathrm{v} + m_\mathrm{v}^2$, where $E$ denotes mathematical expectation.
Therefore, if $\mathbf{v}[k]\sim \mathcal{N}(0, s^2_\mathrm{v})$, then $E \sigma^2_\mathrm{v}[k] = s^2_\mathrm{v}$, and $\tilde{\mathbf{v}}[k]$ is an empirical approximation of WGN$\sim \mathcal{N}(0, 1)$. 
\end{enumerate}
\end{theorem}
\paragraph{Proof} see \ref{appC}. 

\section{Experimental Results}\label{sec5}
In the following sections, we provide synthetic and real data examples demonstrating the performance of the proposed technique.

\subsection {Natural image denoising} \label{sec5a}

We begin with image denoising for the classic case of additive white Gaussian noise (AWGN). As image denoising in this case is, to some extent, considered a solved problem \cite{Elad:2023}, the purpose of this example is mainly to perform a ``sanity check", showing that our proposed strategy can indeed improve denoising in the classic case. 
To this end, we used the dataset presented in \cite{Romano:2017,Cohen:2021}, which consists of 10 common grayscale and color images, referred to as Set10.
The images are contaminated with AWGN of noise level of $\sigma_\mathrm{w}=10,15,25$. Pixel values are in the range $[0,255]$. Color images were converted to YcBcR domain, where each channel is processed separately, and then converted back to RGB. 
The denoising engine we use is the state-of-the-art trainable nonlinear reaction diffusion (TNRD) \cite{Chen:2016} method. This algorithm trains a nonlinear reaction-diffusion model in a supervised manner. In the experiments below we built upon the published pretrained model by the authors of TNRD, tailored to denoise images that are contaminated
by AWGN with a fixed noise level, $\sigma_{\mathrm{w}}=5$. 
Tables \ref{Table1}-\ref{Table3} present PSNR scores for each recovered image for TNRD \cite{Chen:2016}, BM3D \cite{Dabov:2006}, RED-SD \cite{Romano:2017}, and BTB, and $T_{\mathrm{BTB}}$ - the number of iterations required for BTB. BM3D is considered one of the very best classical approaches for image denoising in
terms of MSE results. It relies on the sparsity of 2D-DCT transform of local patches and similarity between image patches.
Note that BM3D parameters are determined by a known noise level. 

\begin{table}[h!]
\centering
\caption{PSNR (dB) scores for natural image denoising, $\sigma_\mathrm{w}=10$, input PSNR=28.10dB}
\label{Table1}
{\scriptsize
  \begin{tabular}{|m{3.2em}|m{3.4em}|m{2.2em}|m{2.3em}|m{2.2em}|m{2.3em}|m{2.0em}|m{2.8em}|m{2.7em}|m{2.7em}|m{2.4em}|m{2.8em}|} %|c|c|c|c|c|c|c|c|c|c|c|
    \hline
							             &Butterfly &Boats  &C.Man   &House   &Parrot  &Lena   &Barbara  &Starfish &Peppers &Leaves  &Average   \\ \hline \hline
									 
		BM3D 		   &35.36  	  &35.09  &34.18  &36.71  &37.36   &35.21  &34.78    &35.21    &34.68   &35.35  &35.39    \\  \hline
		
		TNRD			 &33.52	    &29.50  &29.38  &29.55  &34.64   &29.51  &29.86    &33.59    &29.36   &33.58  &31.25  \\	\hline
		
RED:SD-TNRD  	 &30.55  	  &30.05  &30.07  &30.15  &30.52   &30.15  &29.39    &30.56    &29.78   &30.72  &30.19  \\  \hline
		\rowcolor{Gray}
    BTB 			 &34.03   	&34.89	&34.13  &35.93  &35.68	 &35.27  &34.12    &34.09    &33.55   &33.93  &34.56 \\  \hline
		
		$T_{\mathrm{BTB}}$        &3         &6      &6      &6       &3  	   &6	     &7        &3         &3      &3    &4.6    \\ %

			\hline 
  \end{tabular} 
	}
\end{table}

\begin{table}[h!]
\centering
\caption{PSNR (dB) scores for natural image denoising, $\sigma_\mathrm{w}=15$, input PSNR=24.58dB}
\label{Table2}
{\scriptsize
  \begin{tabular}{|m{3.2em}|m{3.4em}|m{2.2em}|m{2.3em}|m{2.2em}|m{2.3em}|m{2.0em}|m{2.8em}|m{2.7em}|m{2.7em}|m{2.4em}|m{2.8em}|} %|c|c|c|c|c|c|c|c|c|c|c|
    \hline
							             &Butterfly &Boats  &C.Man   &House   &Parrot  &Lena   &Barbara  &Starfish &Peppers &Leaves  &Average   \\ \hline \hline
									 
		BM3D 		   &33.05  	  &32.93  &31.92  &34.94  &35.43   &33.04  &32.73    &32.96    &32.31   &33.13  &33.24  \\  \hline
		
		TNRD			 &26.88	    &25.21  &25.23  &25.27  &27.15   &25.24  &25.19    &27.09    &25.20   &27.64  &26.01 \\	\hline
		
RED:SD-TNRD  	 &31.10  	  &26.63  &26.71  &26.74  &32.42   &26.64  &26.44   &31.50    &26.53   &31.10   &28.58 \\  \hline
		\rowcolor{Gray}
    BTB 			 &32.67   	&32.77	&31.77  &34.29  &34.26	 &32.94  &31.71    &32.63    &31.70   &32.34  &32.70  \\  \hline
		
		$T_{\mathrm{BTB}}$        &4         &12      &12     &13     &4  	   &12	    &13       &4        &12      &4 &9        \\ %

			\hline 
  \end{tabular} 
	}
\end{table}

\begin{table}[h!]
\centering
\caption{PSNR (dB) scores for natural image denoising, $\sigma_\mathrm{w}=25$, input PSNR=20.14dB}
\label{Table3}
{\scriptsize
 \begin{tabular}{|m{3.2em}|m{3.4em}|m{2.2em}|m{2.3em}|m{2.2em}|m{2.3em}|m{2.0em}|m{2.8em}|m{2.7em}|m{2.7em}|m{2.4em}|m{2.8em}|} %|c|c|c|c|c|c|c|c|c|c|c|
    \hline
							             &Butterfly &Boats  &C.Man   &House   &Parrot  &Lena   &Barbara  &Starfish &Peppers &Leaves  &Average   \\ \hline \hline
									 
		BM3D 		   &30.21  	  &30.26  &29.45  &32.86  &32.84   &30.39  &30.03    &30.27    &30.16   &30.26  &30.67  \\  \hline
		
		TNRD			 &21.33	    &20.34  &20.25  &20.25  &21.43   &20.35  &20.35    &21.51    &20.34   &22.08  &20.82  \\	\hline
		
RED:SD-TNRD  	 &29.59 	  &21.61  &22.02  &21.64  &30.66   &26.64  &26.44    &29.30    &21.68   &28.49  &25.81 \\  \hline
		\rowcolor{Gray}
    BTB 			 &29.87   	&28.99	&28.59  &31.67  &30.60	 &29.50  &28.02    &29.58    &28.74  &28.74   &29.43 \\  \hline
		
		$T_{\mathrm{BTB}}$        &10        &37     &37     &40     &14  	   &37    &36        &10       &38      &9   &26.80     \\ %

			\hline 
  \end{tabular} 
	}
\end{table}

\begin{figure}[t!]
\centering
    \begin{subfigure}[t]{0.16\textwidth}
        \includegraphics[width=0.9\linewidth]{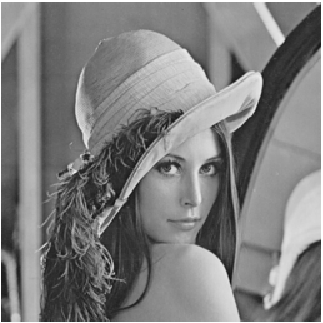}
				\caption{}
    \end{subfigure}
		\begin{subfigure}[t]{0.16\textwidth}
        \includegraphics[width=0.9\linewidth]{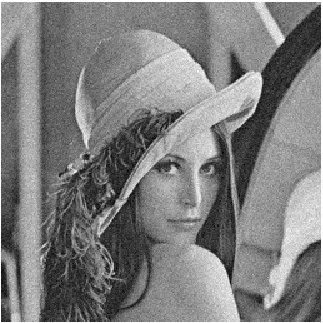}
				\caption{}
    \end{subfigure}
    \begin{subfigure}[t]{0.16\textwidth}
        \includegraphics[width=0.9\linewidth]{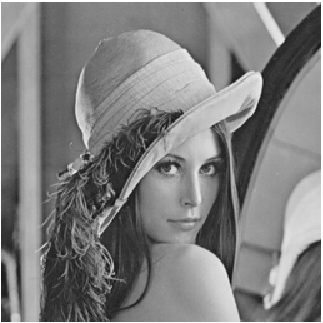}
				\caption{}
    \end{subfigure}
		\begin{subfigure}[t]{0.16\textwidth}
        \includegraphics[width=0.9\linewidth]{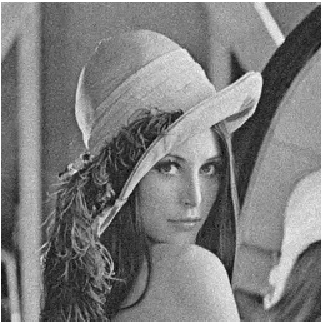}
				\caption{}
    \end{subfigure}
		\begin{subfigure}[t]{0.16\textwidth}
        \includegraphics[width=0.9\linewidth]{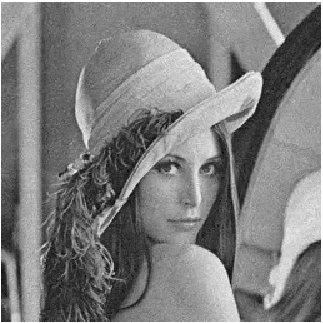}
				\caption{}
    \end{subfigure}
		\begin{subfigure}[t]{0.16\textwidth}
        \includegraphics[width=0.9\linewidth]{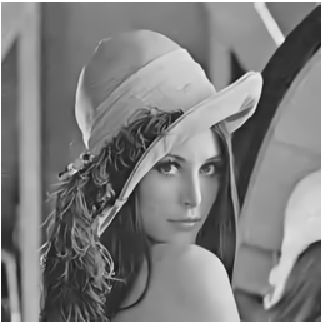}
				\caption{}
    \end{subfigure}
\caption{{\footnotesize Visual comparison of denoising of the image Lena, AWGN with $\sigma_w=10$: (a) Ground truth; (b) input, 28.10dB; (c) BM3D, 35.21dB; (d) TNRD 29.51dB; (e) RED-SD, 30.15dB; (f) BTB, 35.17dB.}}
\label{fig3} %
\end{figure}

\begin{figure}[t!]
\centering
    \begin{subfigure}[t]{0.16\textwidth}
        \includegraphics[width=0.9\linewidth]{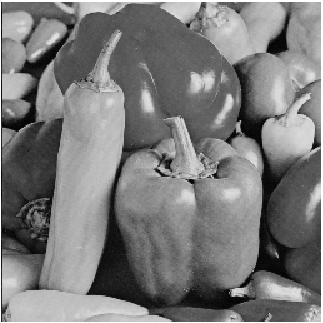}
				\caption{}
    \end{subfigure}
		\begin{subfigure}[t]{0.16\textwidth}
        \includegraphics[width=0.9\linewidth]{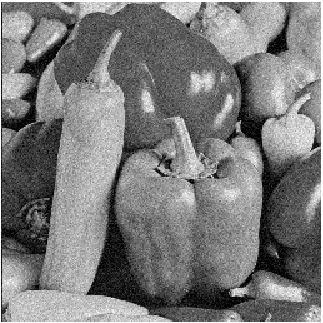}
				\caption{}
    \end{subfigure}
    \begin{subfigure}[t]{0.16\textwidth}
        \includegraphics[width=0.9\linewidth]{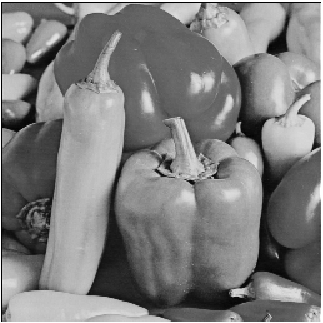}
				\caption{}
    \end{subfigure}
		\begin{subfigure}[t]{0.16\textwidth}
        \includegraphics[width=0.9\linewidth]{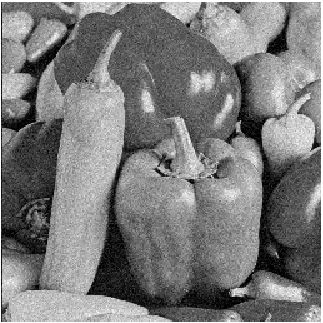}
				\caption{}
    \end{subfigure}
		\begin{subfigure}[t]{0.16\textwidth}
        \includegraphics[width=0.9\linewidth]{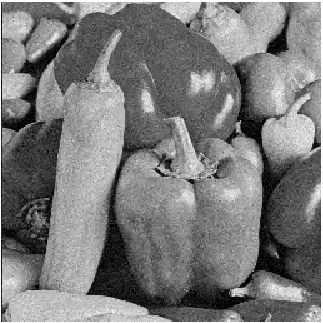}
				\caption{}
    \end{subfigure}
		\begin{subfigure}[t]{0.16\textwidth}
        \includegraphics[width=0.9\linewidth]{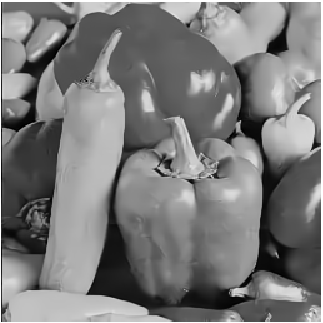}
				\caption{}
    \end{subfigure}
\caption{{\footnotesize Visual comparison of denoising of the image Peppers, AWGN with $\sigma_w=15$: (a) Ground truth; (b) input, 20.18dB; (c) BM3D, 32.70dB; (d) TNRD 25.19dB; (e) RED-SD, 26.53dB; (f) BTB, 31.70dB.}}
\label{fig4} %
\end{figure}

\begin{figure}[th!]
\centering
    \begin{subfigure}[t]{0.16\textwidth}
        \includegraphics[width=0.9\linewidth]{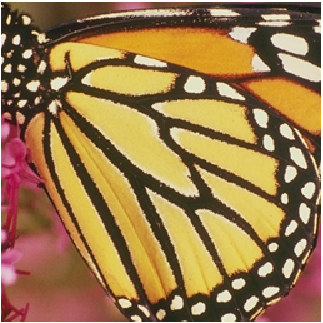}
				\caption{}
    \end{subfigure}
		\begin{subfigure}[t]{0.16\textwidth}
        \includegraphics[width=0.9\linewidth]{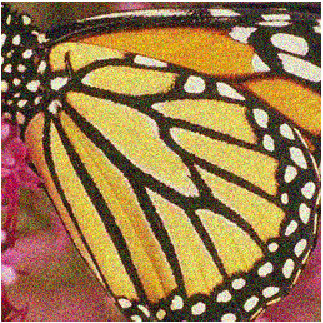}
				\caption{}
    \end{subfigure}
    \begin{subfigure}[t]{0.16\textwidth}
        \includegraphics[width=0.9\linewidth]{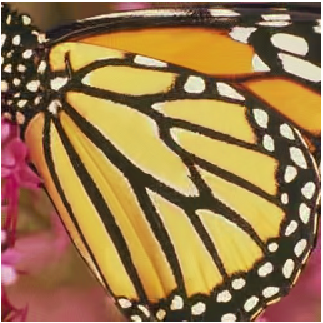}
				\caption{}
    \end{subfigure}
		\begin{subfigure}[t]{0.16\textwidth}
        \includegraphics[width=0.9\linewidth]{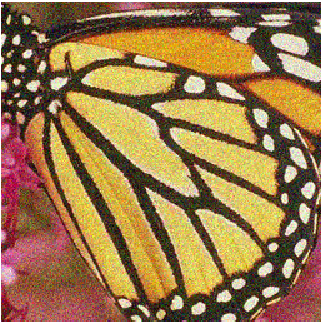}
				\caption{}
    \end{subfigure}
		\begin{subfigure}[t]{0.16\textwidth}
        \includegraphics[width=0.9\linewidth]{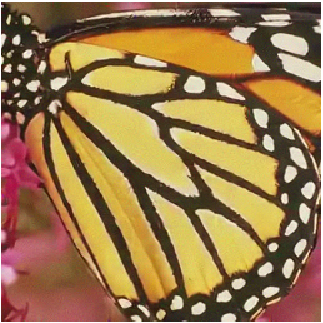}
				\caption{}
    \end{subfigure}
		\begin{subfigure}[t]{0.16\textwidth}
        \includegraphics[width=0.9\linewidth]{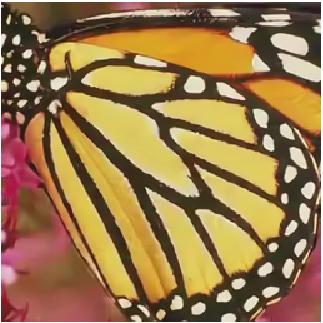}
				\caption{}
    \end{subfigure}
\caption{{\footnotesize Visual comparison of denoising of the image Butterfly, AWGN with $\sigma_w=25$: (a) Ground truth; (b) input, 20.18dB; (c) BM3D, 30.21dB; (d) TNRD 21.33dB; (e) RED-SD, 29.59dB; (f) BTB, 29.87dB.}}
\label{fig5} %
\end{figure}

Figures \ref{fig3}-\ref{fig5} present visual comparison for 3 example images. As can be seen, our denoising method achieves clear and real-looking results. Averaging artifacts are visible as the noise level increases, depending on the denoiser employed. Note that, as stated above, the noise after the first BTB iteration is no longer white noise as typically assumed, yet the BTB denoiser is still able to significantly improve the final output PSNR. Indeed, in terms of PSNR score, BM3D outperforms BTB by a relatively small difference in almost all cases. That said, it is known that PSNR scores, reflecting the MSE, have known drawbacks \cite{Wang:2009}. Hence, we still see the merits in reporting these results. Blau et al.(2018) \cite{Blau:2018} defined the perception-distortion trade-off, stating that a prediction that minimizes a mean distance in any metric will necessarily suffer from a degradation in perceptual quality. It was also proven in \cite{Blau:2018}
that perfect perceptual quality can be obtained without sacrificing more than a factor
of 2 in MSE (3dB in PSNR). That could serve as a good baseline for evaluation of our results.

Our results reassures us of the validity of the proposed approach, namely, ``dropping" the fidelity term can facilitate denoising even in the basic case of AWGN. 
Furthermore, the proposed approach can also practically serve, given access to a trained denoiser with unknown or a mismatch of noise level (known as blind denoising). 
Moreover, we observed that the proposed iterations indeed stably converge, and the denoiser gradually decreases the noise's energy, reassuring us of the validity of Theorems~\ref{Theorem1}-\ref{Theorem2}.   

We observe that color images tend to have higher RED score, since we are denoising each channel separately and by adding them back together, we are averaging 3 channels which could potentially lead to a PSNR increase of $\sim$4.77dB.  
(RED was not designed initially for denoising over 3 channels in this manner by the authors of \cite{Romano:2017}).

Table~\ref{Table6} compares PSNR and SSIM average scores BTB with BM3D and TNRD over a test dataset containing 68 natural
images from Berkeley segmentation dataset (BSD68) \cite{BSD68:2001}, and indicates the number of iterations required for BTB. 
Set10 and BSD68 consist of images that are widely used for the evaluation of denoising methods.

\begin{table}[h!]
\centering
\caption{Gaussian denoising average PSNR (dB) / SSIM scores for BSD68}
\label{Table6}
{\scriptsize
  \begin{tabular}{|m{4.8em}|m{5.8em}|m{5.8em}|m{5.8em}|m{5.8em}|m{3.8em}|} %|c|c|c|c|c|c|c|c|c|c|c|
    \hline
		Noise level											&Input 		     & BM3D   			  &TNRD 	  		    &BTB  	 		        & $T_{\mathrm{BTB}}$   \\ \hline \hline
									 
		$\sigma_{\mathrm{w}}=10$	   		&28.13/0.7072	 &33.32/0.9158  	&29.30/0.7538     &33.38/0.9216      & 5  							\\  \hline
		
		$\sigma_{\mathrm{w}}=15$				&24.62/0.5682  &31.08/0.8722    &25.21/0.5929     &30.94/0.8673		   & 12     					\\	\hline
		
		$\sigma_{\mathrm{w}}=25$				&20.17/0.3095	 &28.57/0.8017    &28.92/0.3969  		&27.80/0.7714      & 36   			      \\  \hline
		
  \end{tabular} 
	}
\end{table}

\subsection{Poisson Noise for Natural Image Denoising}
Photon noise in optical imaging is typically modeled as a Poisson process \cite{Ratner:2007}.
%A pixel is characterized by its full well capacity (FWC), that
%is, the maximal amount of charge that can be accumulated
%in the pixel before saturating. Let $ x \in [0, FWC]$ be the
%expected (noiseless) value of a measured signal (photoelectrons). The corresponding noisy measurement is
That is, a pixel's value is described as $y=\mathrm{Poisson}(x)$ \cite{Torem:2023}.
Here, the variance increases with the expected measured intensities. 
To simplify its analysis, Poisson noise is sometimes approximated by a Gaussian distribution.
Namely,
\begin{equation}
y = x + w, \qquad w\sim\mathcal{N}(0,x).
\end{equation}
The noise variance is linear in $x$.

\begin{figure}[th!]
\centering
    \begin{subfigure}[t]{0.19\textwidth}
        \includegraphics[width=0.9\linewidth]{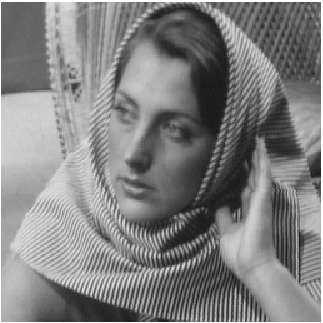}
				\caption{}
    \end{subfigure}
		\begin{subfigure}[t]{0.19\textwidth}
        \includegraphics[width=0.9\linewidth]{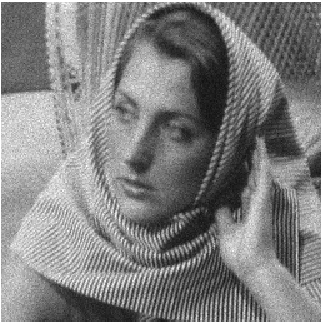}
				\caption{}
    \end{subfigure}
    \begin{subfigure}[t]{0.19\textwidth}
        \includegraphics[width=0.9\linewidth]{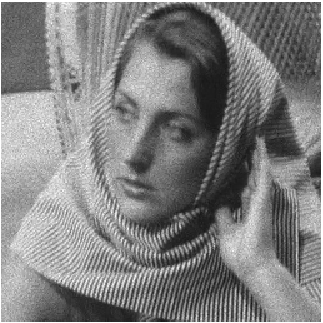}
				\caption{}
    \end{subfigure}
		\begin{subfigure}[t]{0.19\textwidth}
        \includegraphics[width=0.9\linewidth]{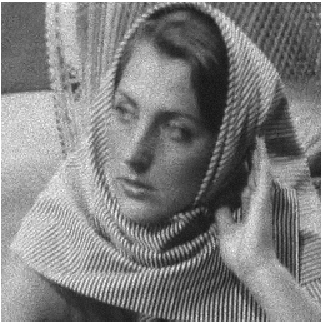}
				\caption{}
    \end{subfigure}
		\begin{subfigure}[t]{0.19\textwidth}
        \includegraphics[width=0.9\linewidth]{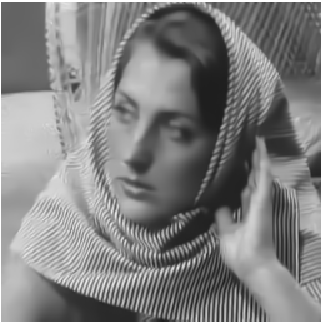}
				\caption{}
    \end{subfigure}
\caption{{\footnotesize Visual comparison of denoising of the image Barbara, Poisson noise: (a) Ground truth; (b) input, 27.07dB; 
(c) TNRD 28.09dB; (e) RED-SD, 28.31dB; (f) BTB, 33.02dB.}}
\label{fig6}
\end{figure}

\begin{figure}[t!]
\centering
    \begin{subfigure}[t]{0.19\textwidth}
        \includegraphics[width=0.9\linewidth]{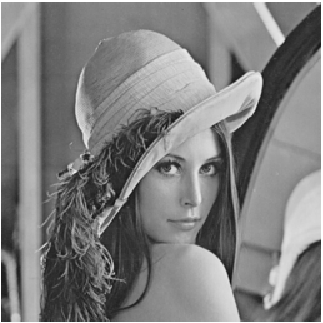}
				\caption{}
    \end{subfigure}
		\begin{subfigure}[t]{0.19\textwidth}
        \includegraphics[width=0.9\linewidth]{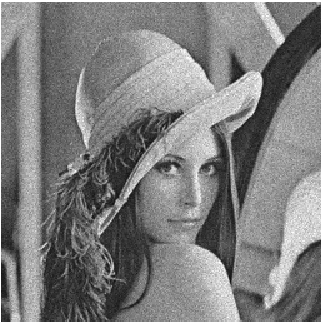}
				\caption{}
    \end{subfigure}
    \begin{subfigure}[t]{0.19\textwidth}
        \includegraphics[width=0.9\linewidth]{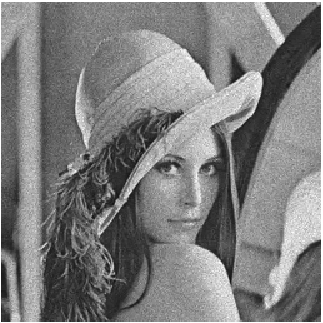}
				\caption{}
    \end{subfigure}
		\begin{subfigure}[t]{0.19\textwidth}
        \includegraphics[width=0.9\linewidth]{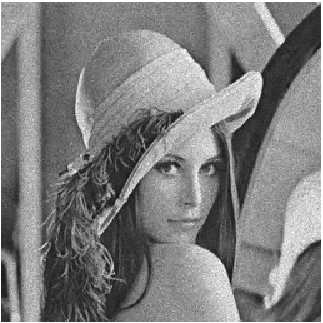}
				\caption{}
    \end{subfigure}
		\begin{subfigure}[t]{0.19\textwidth}
        \includegraphics[width=0.9\linewidth]{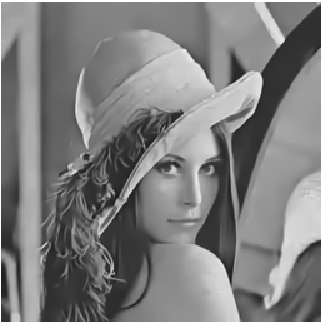}
				\caption{}
    \end{subfigure}
\caption{{\footnotesize Visual comparison of denoising of the image Lena, Poisson noise: (a) Ground truth; (b) input, 27.18dB; 
(c) TNRD 28.31dB; (d) RED-SD, 28.54dB; (e) BTB, 33.95dB.}}
\label{fig7}
\end{figure}

Therefore, we propose using a Gaussian noise denoiser, using the update rule in (\ref{3.1}).
Table \ref{Table4} presents the PSNR values obtained by BTB in comparison with TNRD, and RED-SD (assuming AWGN), for Set10. Figures \ref{fig6}-\ref{fig7} present examples of visual results. As can be seen our method efficiently suppresses noise while preserving structural details, and requires very few iterations. Table~\ref{Table7} compares average PSNR and SSIM scores for BSD68. 

\begin{table}[h!]
\centering
\caption{PSNR (dB) scores for natural image denoising - Poisson noise}
\label{Table4}
{\scriptsize
 \begin{tabular}{|m{3.2em}|m{3.4em}|m{2.2em}|m{2.3em}|m{2.2em}|m{2.3em}|m{2.0em}|m{2.8em}|m{2.7em}|m{2.7em}|m{2.4em}|m{2.8em}|} %|c|c|c|c|c|c|c|c|c|c|c|
    \hline
							             &Butterfly &Boats  &C.Man   &House   &Parrot  &Lena   &Barbara  &Starfish &Peppers &Leaves  &Average   \\ \hline \hline
							
Input PSNR 		 &27.57  	  &26.96  &27.36  &26.70  &27.96   &27.17  &27.08    &28.08    &27.31   &26.39    &27.26\\  \hline
									 
		%BM3D 		   &30.21  	  &30.26  &29.45  &32.86  &32.84   &30.39  &30.03    &30.27    &30.16   &30.26    \\  \hline
		
		TNRD			 &31.12	    &28.02  &28.36  &27.79  &31.86   &28.31  &28.10    &31.86    &28.38   &29.53   &29.33\\	\hline
		
RED:SD-TNRD  	 &30.42 	  &28.18  &28.38  &27.99  &31.19   &28.54  &28.31    &30.19    &28.53   &29.86   &29.16\\  \hline
		\rowcolor{Gray}
    BTB 			 &33.73   	&33.85	&32.92  &34.86  &35.13	 &34.19  &33.01    &33.50    &32.64   &33.17   &33.70\\  \hline
		
		$T$        &3         &8      &9      &9      &3 	     &9      &9        &3        &8       &3       &6.4 \\ %

			\hline 
  \end{tabular} 
	}
\end{table}

\begin{table}[h!]
\centering
\caption{Poisson denoising average PSNR (dB) / SSIM scores for BSD68}
\label{Table7}
{\scriptsize
  \begin{tabular}{|m{5.8em}|m{5.8em}|m{5.8em}|m{5.8em}|m{3.8em}|} %|c|c|c|c|c|c|c|c|c|c|c|
    \hline
		Input 		     &TNRD 	  		    &RED-SD             &BTB  	 		        & $T_{\mathrm{BTB}}$   \\ \hline \hline
									 
		28.04/0.7200	 &29.12/0.7717   &29.33/0.7790				&32.31/0.8937       & 6  							\\  \hline

  \end{tabular} 
	}
\end{table}

\subsection{OCT Speckle Suppression} \label{sec5b}

OCT employs low coherence interferometry to obtain cross-sectional tomographic images of internal structure of biological tissue. It is routinely used for diagnostic imaging, primarily of the retina and coronary arteries \cite{Villiger:2020}. %The resolutions obtainable are in the range 1 to 15 $\mu$m, with depth range of millimeters. 
Unfortunately, OCT images are degraded by speckle \cite{Schmitt:1999,Goodman:2007}, creating visibly prominent apparent grain-like patterns in the image, as large as the spatial resolution of the OCT system. Speckle presence significantly degrades images, obscuring tissue anatomy and changes in tissue scattering properties, which in turn complicates interpretation and thus medical diagnosis. 

In medical imaging, speckle noise has long been an extensively studied problem. 
In recent years, significant progress has been made by employing deep learning methods for noise reduction. However, supervised learning models are still facing challenges in terms of their adaptation to unseen domains. In particular, deep neural networks (DNNs) trained for computational imaging tasks are vulnerable to changes in the acquisition system's physical parameters, such as: sampling space, resolution, and contrast. There is ample evidence of performance issues across datasets of different biological tissues, even within the same acquisition system. Therefore, an alternative unsupervised approach can be useful.

\subsubsection{Synthetic data}

\begin{figure*}[t]
\centering
		\begin{subfigure}[t]{0.16\textwidth}
        \includegraphics[width=\linewidth]{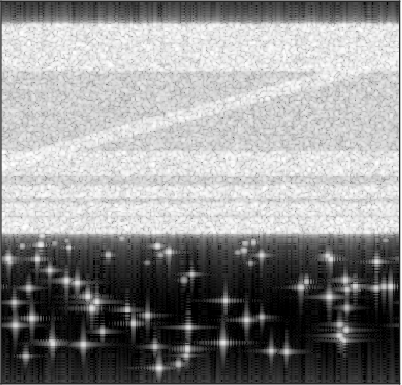}
				\caption*{{\footnotesize $\mathbf{Y}$}}
    \end{subfigure}
		\begin{subfigure}[t]{0.16\textwidth}
        \includegraphics[width=\linewidth]{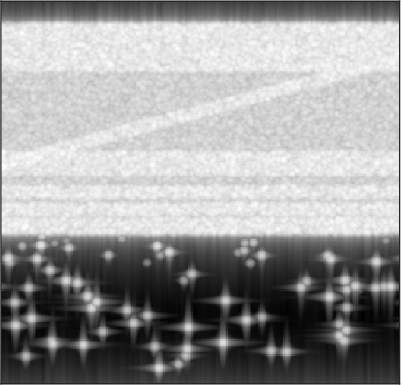}
				\caption*{{\footnotesize $\mathbf{X}$}}
    \end{subfigure} 
				\begin{subfigure}[t]{0.16\textwidth}
        \includegraphics[width=\linewidth]{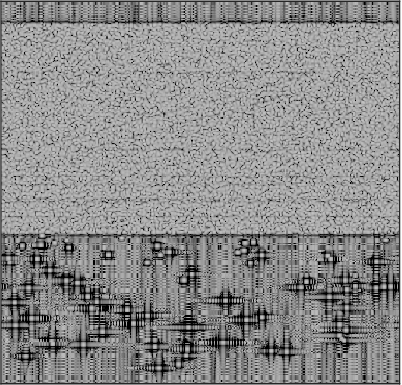}
				\caption*{{\footnotesize $\mathbf{Y}-\mathbf{X}$}}
    \end{subfigure}%
		\\
    \begin{subfigure}[t]{0.16\textwidth}
        \includegraphics[width=\linewidth]{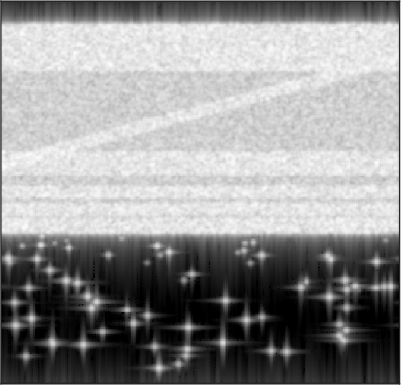}
				\caption*{{\footnotesize $t=1$}}
    \end{subfigure}
		\begin{subfigure}[t]{0.16\textwidth}
        \includegraphics[width=\linewidth]{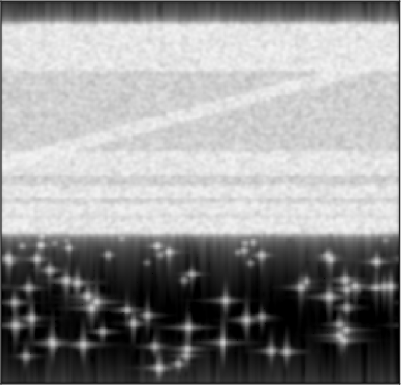}
				\caption*{{\footnotesize $t=3$}}
    \end{subfigure}
		\begin{subfigure}[t]{0.16\textwidth}
        \includegraphics[width=\linewidth]{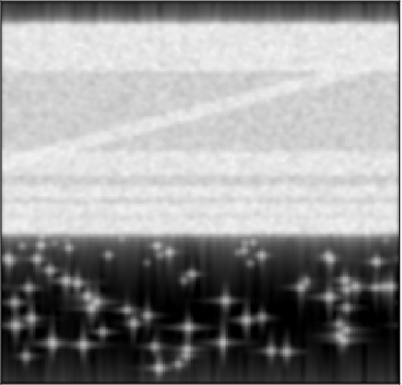}
				\caption*{{\footnotesize $t=5$}}
    \end{subfigure}
		\begin{subfigure}[t]{0.19\textwidth}
        \includegraphics[width=\linewidth,height=2.05cm]{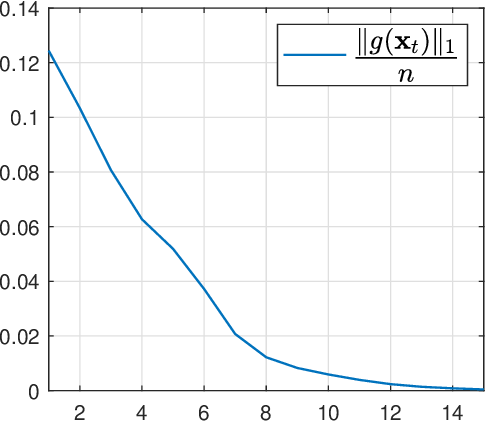}
				\caption*{{\scriptsize $t$}}
    \end{subfigure}
		\caption{Visualization of speckle suppression synthetic example. Left-to-right. First row: Speckled tomogram, speckle-free incoherent mean (ground truth), residual (noise) image. Second row: Vortice outputs, iterations: $t=\{1,3,5\}$;  Evolution of the degree of speckle.}
\label{fig9}
\end{figure*}

To test the validity of our approach, we built a synthetic example of layered structure.
The image was generated in Matlab \cite{MATLAB:2021}, assuming wavelength-swept light source with a $\lambda_c = 1300 nm$ central wavelength, and a spectral bandwidth of $\Delta \lambda = 130 nm$. The distance between A-lines was $5\mu m$, generating images of size $512\times 512$.
Particle densities were in the range $[0.0005,0.5]$. For all experiments we set $\alpha=0.4$.

Let us denote $f(\mathscr{z,x}) \in \mathbb{C}$ as the ground truth ideal tomogram perfectly describing the depth sample reflectivity.
Here $\big\{(\mathscr{z,x}): \mathscr{z,x} \geq 0,  (\mathscr{z,x})\in \mathbb{R}^2 \big\}$ are continuous axial and lateral spatial axes. A measured tomogram can be formulated as 
\begin{equation}\label{5.1}
Y(\mathscr{z,x})=10\log_{10}\big(|f(\mathscr{z,x})*\alpha(\mathscr{z,x})|^2\big).
\end{equation}
where $*$ denotes the convolution operation and $\alpha(z,x)$ is a point spread function (PSF).
In the discrete setting, assuming $F^{z}_{\mathrm{s}}$, $F^{x}_{\mathrm{s}}$ axial and lateral sampling rates respectively, and that the set of measured values at $\{\mathscr{z}_m,\mathscr{x}_n\}$ lie on the grid $m/F^{z}_{\mathrm{s}}$ and $n/F^{x}_{\mathrm{s}}$, $m,n \in \mathbb{N}$,
the speckled image (Fig.~\ref{fig9}) is modeled as 
\begin{equation}\label{5.2}
Y[m,n]=10\log_{10}\big(\big|f[m,n]*\alpha[m,n]\big|^2\big)
\end{equation}
A speckle-free tomogram can be viewed as the incoherent mean of coherent tomograms with different speckle realizations \cite{Goodman:2007}, namely %,van:2012},
\begin{equation}\label{5.3}
X[m,n]=10\log_{10}\big(\big|f[m,n]\big|^2*\big|\alpha[m,n]\big|^2\big)
\end{equation}

Figure~\ref{fig9} presents a speckled tomogram, its corresponding ground truth, the residual signal and the estimation $\mathbf{x}_t$ at iterations $t=\{1,3,5\}$ of Algorithm~\ref{alg4}. As can be seen, the algorithm effectively suppresses speckle already in the first iteration. To evaluate the correspondence between the RFN operator to the degree of speckle presence in the image, we plot $\|g(\mathbf{x}_t)\|_1/n$ as the number of iterations progresses. We observe that this average norm term consistently decreases at each iteration, approaching zero, clearly indicating that the RFN denoiser converges to a fixed point solution. Hence the RFN operator output norm converges, and it can be used as a reliable estimator of the speckle-level in the image.  
Note that, as in other ill-posed reconstruction problems, the observed speckled image may originate in many plausible reconstructions with varying textures and ﬁne details, and different semantic information \cite{Kawar:2021}.

%(referred to as Vortice)

\subsubsection{Real Data}

\begin{figure*}[t]
				    \begin{subfigure}[t]{0.14\textwidth}
        \includegraphics[width=\linewidth]{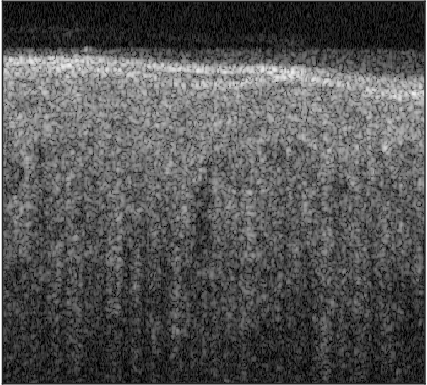}
    \end{subfigure}%
		\hfill 
		\begin{subfigure}[t]{0.14\textwidth}
        \includegraphics[width=\linewidth]{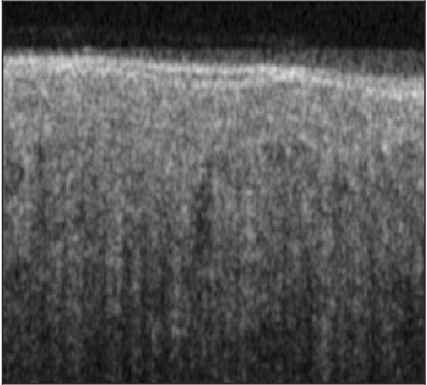}
    \end{subfigure}%
		\hfill 
    \begin{subfigure}[t]{0.14\textwidth}
        \includegraphics[width=\linewidth]{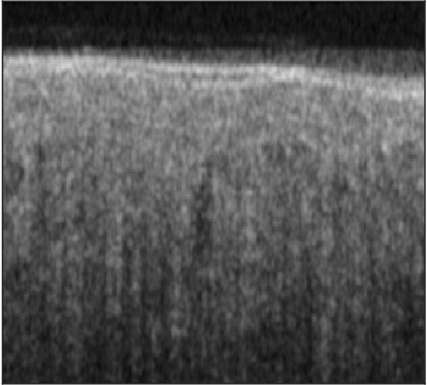}
    \end{subfigure}%
		\hfill 
		\begin{subfigure}[t]{0.14\textwidth}
        \includegraphics[width=\linewidth]{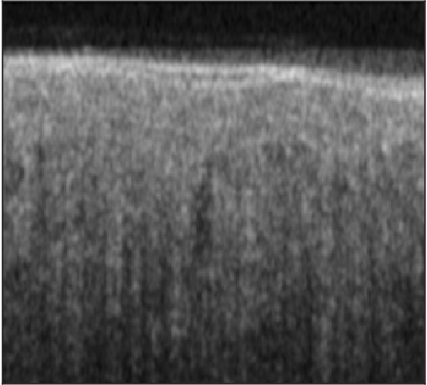}
    \end{subfigure}%
				\hfill 
		\begin{subfigure}[t]{0.14\textwidth}
        \includegraphics[width=\linewidth]{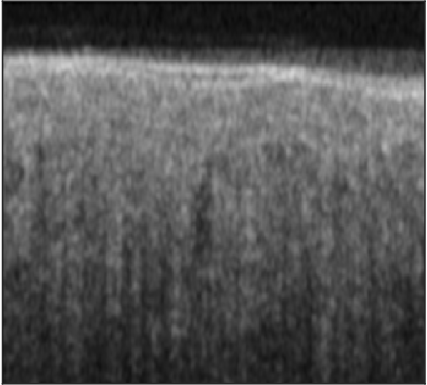}
    \end{subfigure}%
				\hfill 
		\begin{subfigure}[t]{0.14\textwidth}
        \includegraphics[width=\linewidth]{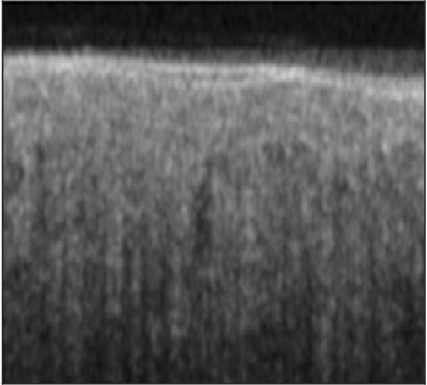}
    \end{subfigure}%
				\hfill 
		\begin{subfigure}[t]{0.14\textwidth}
        \includegraphics[width=\linewidth]{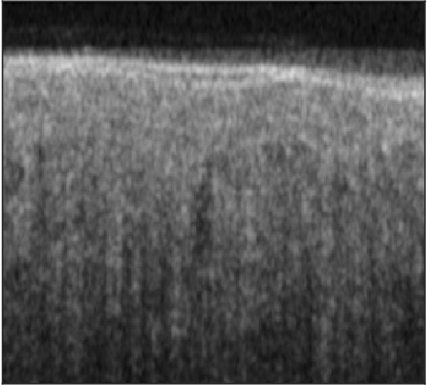}
    \end{subfigure}

				    \begin{subfigure}[t]{0.14\textwidth}
        \includegraphics[width=\linewidth]{fig3_S.eps}
    \end{subfigure}%
		\hfill 
		\begin{subfigure}[t]{0.14\textwidth}
        \includegraphics[width=\linewidth]{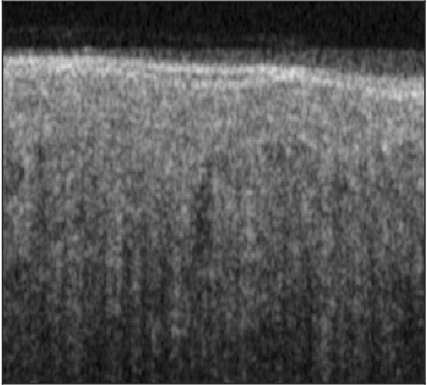}
    \end{subfigure}%
		\hfill 
    \begin{subfigure}[t]{0.14\textwidth}
        \includegraphics[width=\linewidth]{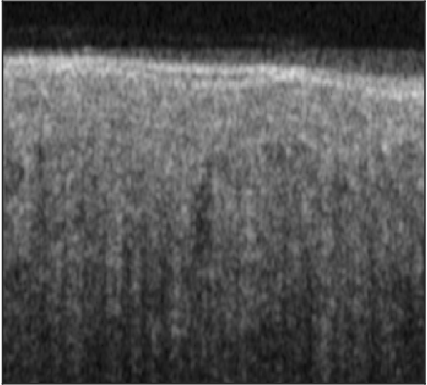}
    \end{subfigure}%
		\hfill 
		\begin{subfigure}[t]{0.14\textwidth}
        \includegraphics[width=\linewidth]{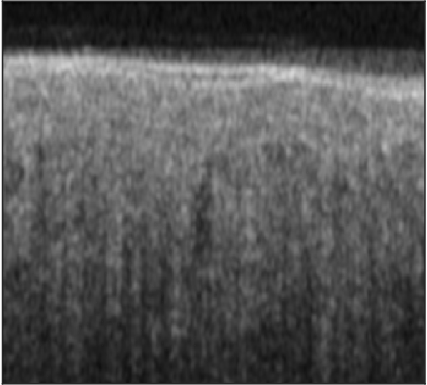}
    \end{subfigure}%
				\hfill 
		\begin{subfigure}[t]{0.14\textwidth}
        \includegraphics[width=\linewidth]{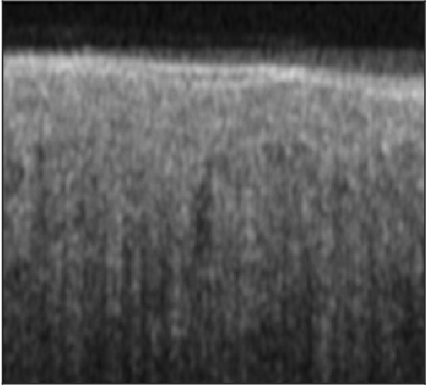}
    \end{subfigure}%
				\hfill 
		\begin{subfigure}[t]{0.14\textwidth}
        \includegraphics[width=\linewidth]{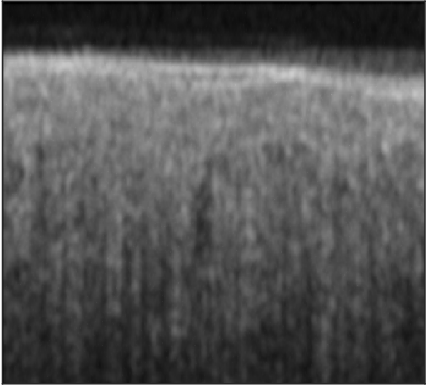}
    \end{subfigure}%
				\hfill 
		\begin{subfigure}[t]{0.14\textwidth}
        \includegraphics[width=\linewidth]{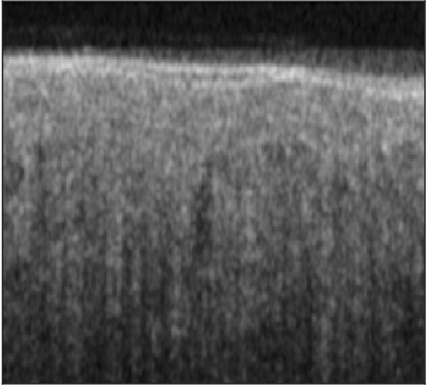}
    \end{subfigure}			

				    \begin{subfigure}[t]{0.14\textwidth}
        \includegraphics[width=\linewidth]{fig3_S.eps}
    \end{subfigure}%
		\hfill 
\begin{subfigure}[t]{0.14\textwidth}
        \includegraphics[width=\linewidth]{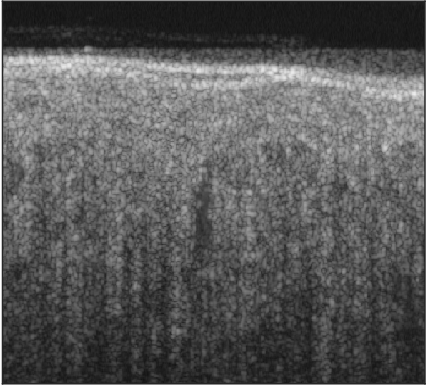}
    \end{subfigure}%
		\hfill 
		\begin{subfigure}[t]{0.14\textwidth}
        \includegraphics[width=\linewidth]{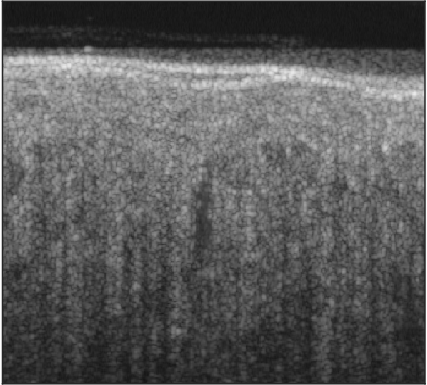}
    \end{subfigure}%
		\hfill 
    \begin{subfigure}[t]{0.14\textwidth}
        \includegraphics[width=\linewidth]{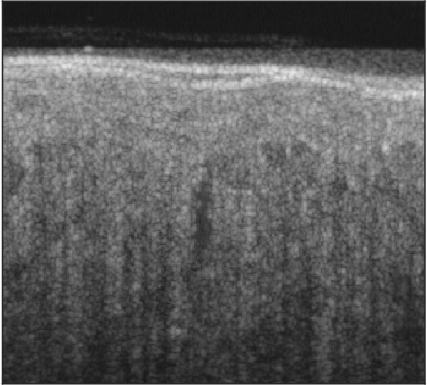}
    \end{subfigure}%
		\hfill 
		\begin{subfigure}[t]{0.14\textwidth}
        \includegraphics[width=\linewidth]{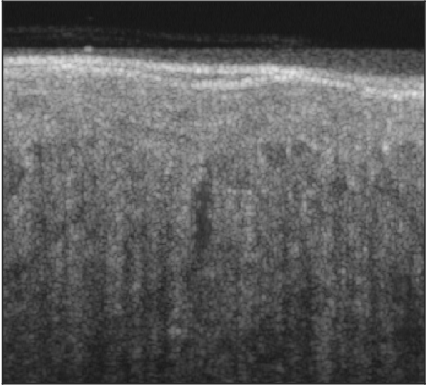}
    \end{subfigure}%
				\hfill 
		\begin{subfigure}[t]{0.14\textwidth}
        \includegraphics[width=\linewidth]{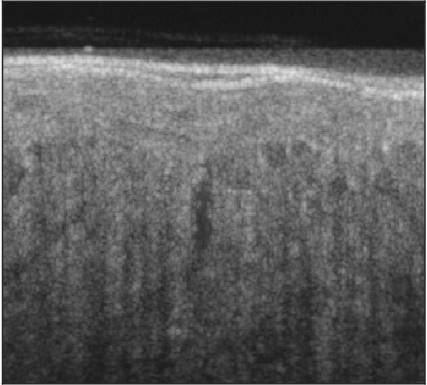}
    \end{subfigure}%
				\hfill 
		\begin{subfigure}[t]{0.14\textwidth}
        \includegraphics[width=\linewidth]{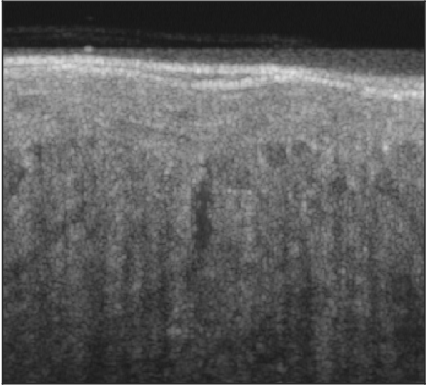}
    \end{subfigure}%
\caption{Visualization of chicken muscle speckle suppression cross-sectional OCT image.
First column (left) - measured speckled-image. 
The first and the second row show a sequence of
images $\mathbf{x}_t,t=\{1,2,...,5\}$ of Algorithm \ref{alg2} and Algorithm \ref{alg4}, respectively. 
Right  - last column - shows mean image averaged over the rest of the images in the same row. Input: PSNR 25.81dB, SSIM 0.23. Algorithm \ref{alg2}: output PSNR 31.64dB, SSIM 0.66. Algorithm \ref{alg4}: output PSNR 31.71dB, SSIM 0.67.
The third row shows angular compounding images of $M=[1,25,45,65,85,105,125]$ tilted stage images \cite{Keahey:2023}. Please zoom-in to observe the details.}
\label{fig10}
\end{figure*}

\begin{figure}[h!]
\centering
    \begin{subfigure}[t]{0.195\textwidth}
        \includegraphics[width=0.9\linewidth]{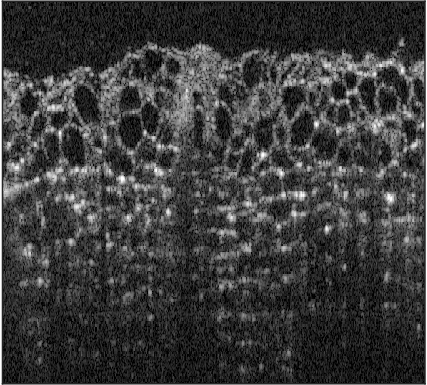}
				\caption{}
    \end{subfigure}%
		\hfill 
		\begin{subfigure}[t]{0.195\textwidth}
        \includegraphics[width=0.9\linewidth]{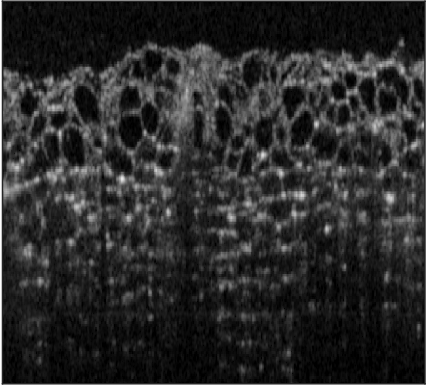}
				\caption{}
    \end{subfigure}%
		\hfill 
    \begin{subfigure}[t]{0.195\textwidth}
        \includegraphics[width=0.9\linewidth]{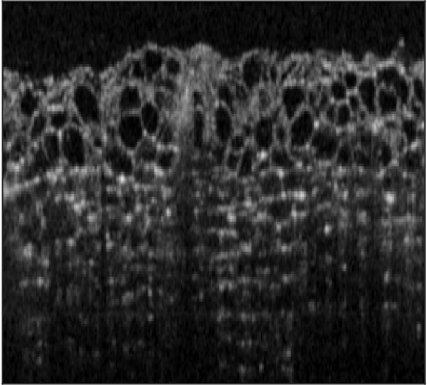}
				\caption{}
    \end{subfigure}%
		\hfill 
		\begin{subfigure}[t]{0.195\textwidth}
        \includegraphics[width=0.9\linewidth]{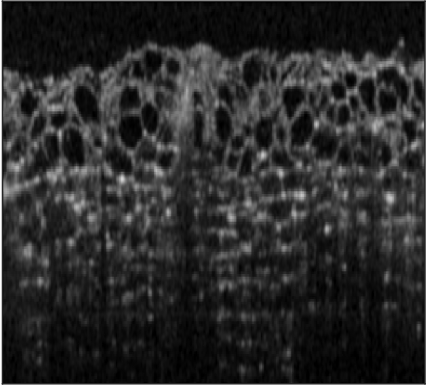}
				\caption{}
    \end{subfigure}%
		\hfill 
		\begin{subfigure}[t]{0.195\textwidth}
        \includegraphics[width=0.9\linewidth]{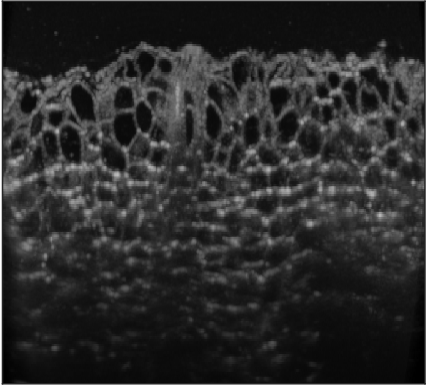}
				\caption{}
    \end{subfigure}%
\caption{{\footnotesize Visual comparison blueberry OCT cross-section speckle suppression: (a) input, PSNR 25.24dB; (b)-(d) $\mathbf{x}_t,t=\{1,2,3\}$ of Algorithm~\ref{alg4}, output PSNR 28.57dB; (e) Ground truth.}}
\label{fig11}
\end{figure}

\paragraph*{Ex vivo OCT samples}
As ground truth for training and testing, we used hardware-based speckle mitigation obtained by dense angular compounding, in a method similar to \cite{Desjardins:2007}.
Ground truth images for chicken muscle and blueberry (Figures~\ref{fig10}-\ref{fig11}), were acquired using an angular compounding (AC) system using sample tilting in combination with a model-based affine transformation to generate speckle suppressed ground truth data \cite{Keahey:2023}. Note that AC via sample tilting is not possible for in vivo samples. 

Figure~\ref{fig10} presents the results of Algorithms \ref{alg2}-\ref{alg4} for 5 iterations, and the averaged image of those iterations, in comparison with acquired OCT tomograms angular compounding processed via increasing number of tilted samples. As can be seen, our approach mimics the compounding process, in a manner that somewhat resembles diffusion models \cite{Sohl:2015,Ho:2020}, where each step gradually improves the image. Table~\ref{Table5} presents quantitative measurements, namely - average PSNR and SSIM scores over 100 images, comparing with few-shot learning approach \cite{Pereg:2023B}.
Visual comparison for blueberry is displayed in Figure~\ref{fig11}.
Intuitively, we are gradually improving image quality. 
That said, we are still averaging solutions, which causes blurring to some extent. 

\paragraph*{Retinal Data}
We used retinal data acquired by a retinal imaging system similar to \cite{braaf:2018}.
As ground truth for training and testing we used NLM-based speckle suppressed images \cite{Cuartas:2018}. 
Note that NLM is considered relatively slow (about 23 seconds for a B-scan of size $1024 \times 1024$). For visualization purposes, images were cropped to size $448 \times 832$.

\begin{figure*}[h!]
				    \begin{subfigure}[t]{0.14\textwidth}
        \includegraphics[width=\linewidth]{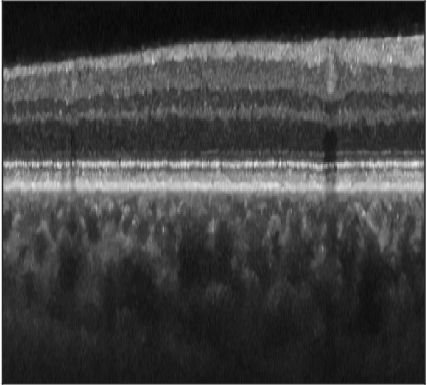}
    \end{subfigure}%
		\hfill 
		\begin{subfigure}[t]{0.14\textwidth}
        \includegraphics[width=\linewidth]{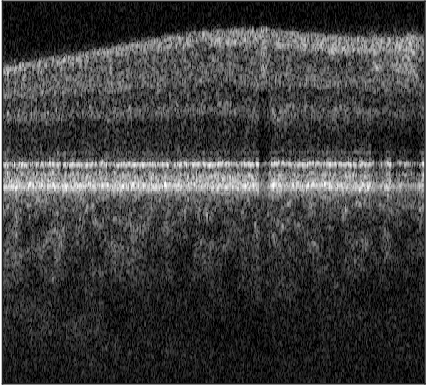}
    \end{subfigure}%
		\hfill 
    \begin{subfigure}[t]{0.14\textwidth}
        \includegraphics[width=\linewidth]{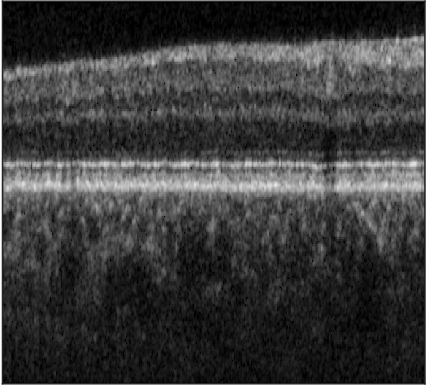}
    \end{subfigure}%
		\hfill 
		\begin{subfigure}[t]{0.14\textwidth}
        \includegraphics[width=\linewidth]{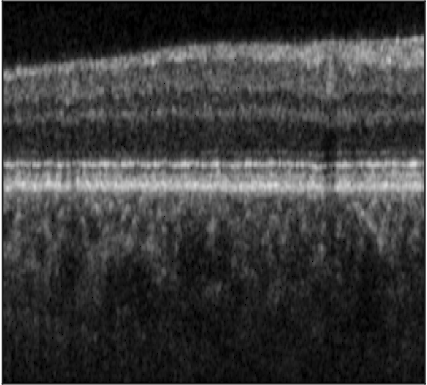}
    \end{subfigure}%
				\hfill 
		\begin{subfigure}[t]{0.14\textwidth}
        \includegraphics[width=\linewidth]{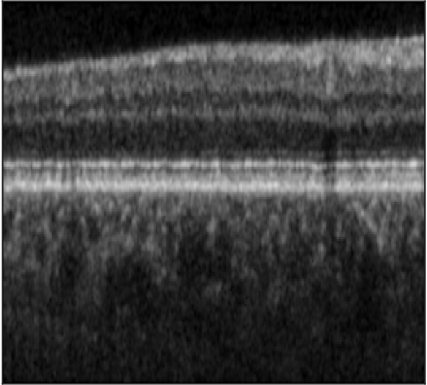}
    \end{subfigure}%
				\hfill 
		\begin{subfigure}[t]{0.14\textwidth}
        \includegraphics[width=\linewidth]{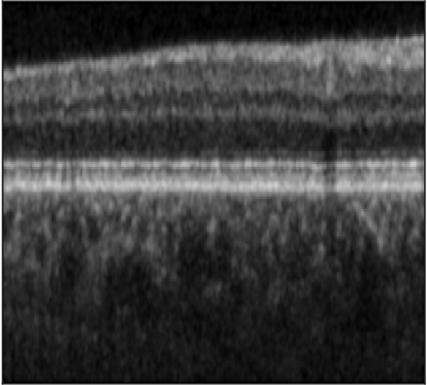}
    \end{subfigure}%
				\hfill 
		\begin{subfigure}[t]{0.14\textwidth}
        \includegraphics[width=\linewidth]{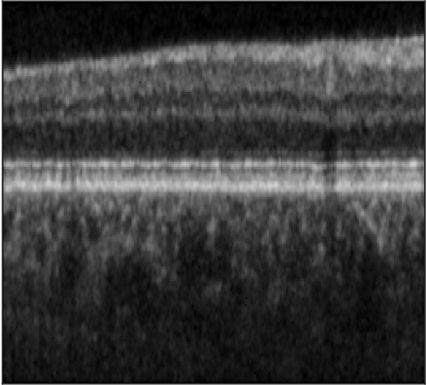}
    \end{subfigure}

				    \begin{subfigure}[t]{0.14\textwidth}
        \includegraphics[width=\linewidth]{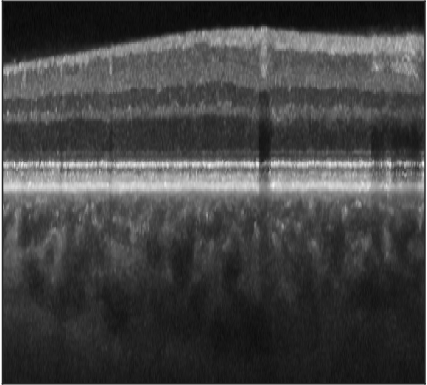}
    \end{subfigure}%
		\hfill 
		\begin{subfigure}[t]{0.14\textwidth}
        \includegraphics[width=\linewidth]{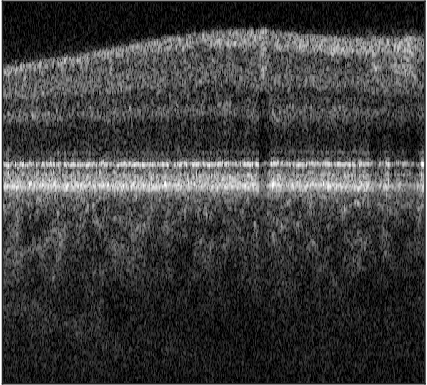}
    \end{subfigure}%
		\hfill 
    \begin{subfigure}[t]{0.14\textwidth}
        \includegraphics[width=\linewidth]{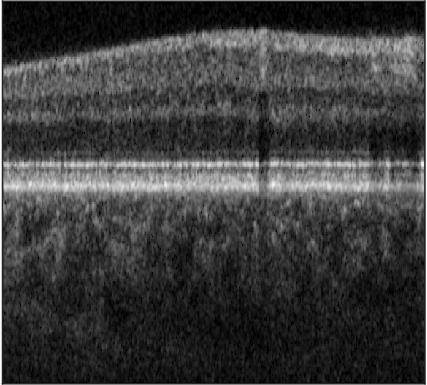}
    \end{subfigure}%
		\hfill 
		\begin{subfigure}[t]{0.14\textwidth}
        \includegraphics[width=\linewidth]{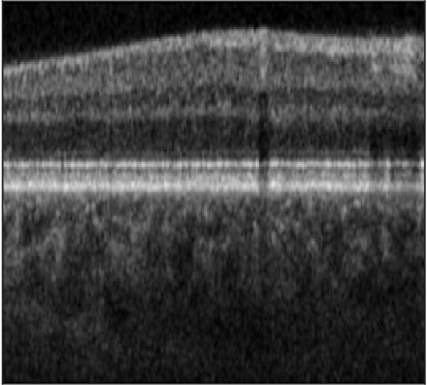}
    \end{subfigure}%
				\hfill 
		\begin{subfigure}[t]{0.14\textwidth}
        \includegraphics[width=\linewidth]{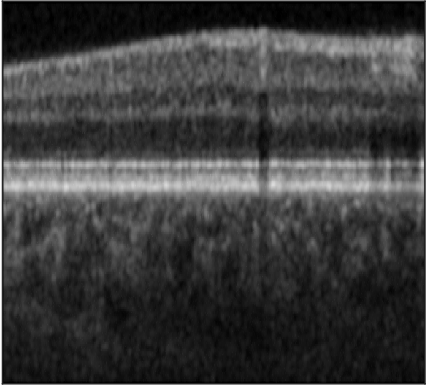}
    \end{subfigure}%
				\hfill 
		\begin{subfigure}[t]{0.14\textwidth}
        \includegraphics[width=\linewidth]{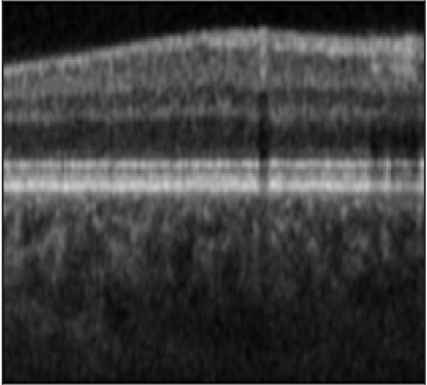}
    \end{subfigure}%
				\hfill 
		\begin{subfigure}[t]{0.14\textwidth}
        \includegraphics[width=\linewidth]{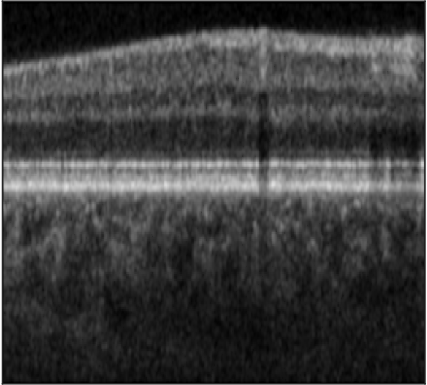}
    \end{subfigure}			
\caption{Visualization of two examples of retinal cross-sectional OCT images speckle suppression:
First column (left) presents NLM results used as ground truth. Second column presents speckled observed image. 
Columns 3-6, show a sequence of
images $\mathbf{x}_t,t=\{2,4,6,8\}$ of Algorithm \ref{alg2} (first row) and Algorithm \ref{alg4} (second row). 
Right last column - shows mean image averaged over estimates of iterations 1-9. 
First row - Input: PSNR 23.25dB, SSIM 0.46. Algorithm \ref{alg2}: output PSNR 28.22dB, SSIM 0.83.
Second row - Input: PSNR 23.40dB, SSIM 0.45. Algorithm \ref{alg4}: output PSNR 29.12dB, SSIM 0.83.}
\label{fig12}
\end{figure*}

Figure~\ref{fig12} presents two examples of retinal cross-sections, and their denoised version over iterations $t=\{2,4,6,8\}$ of Algorithms \ref{alg2}-\ref{alg4}, in comparison with the corresponding NLM despeckled image. As can bee seen, our method, gradually removes the level of speckle, while preserving structural information. Notably, our results are less ``washed out" comparing with NLM, and the original texture is preserved. %Note that NLM is considered relatively slow (about 23 seconds for a B-scan of size $1024 \times 1024$).

%$256 \times 256$	

\begin{table}[h!]
\centering
\caption{Average PSNR (dB) and SSIM scores for OCT despckling}
\label{Table5}
{\scriptsize
  \begin{tabular}{|l|ccc|ccc|} %|c|c|c|c|c|c|c|c|c|c|c|
    \hline
		&\multicolumn{3}{c|}{PSNR (dB)} & \multicolumn{3}{c|}{SSIM} \\ 
	             &Input & Vortice &RNN-GAN             & Input & Vortice & RNN-GAN      \\ \hline \hline
							
Retina 	 	&23.33 $\pm$ 0.08	  &29.02 $\pm$ 0.17    &32.34 $\pm$ 0.12 
													&0.46  $\pm$ 0.01  &0.83  $\pm$ 0.02    &0.86  $\pm$ 0.04     \\	\hline
Chicken 	&24.29 $\pm$ 0.28	  &29.14 $\pm$ 0.21    &30.80 $\pm$ 0.16 
																	&0.28  $\pm$ 0.01  &0.68 $\pm$ 0.02   &0.74  $\pm$ 0.01     \\	\hline
																	
Blueberry 	&25.11 $\pm$ 0.19	  &27.16 $\pm$ 0.12    &28.18 $\pm$ 0.01 
														&0.48  $\pm$ 0.01   &0.67 $\pm$ 0.01      &0.76  $\pm$ 0.01     \\	\hline
		%
%Blueberry  	 &30.42 	  &28.18  &28.38  &27.99  &31.19   &28.54  &28.31    &30.19    &28.53   &29.86    \\  \hline
		%\rowcolor{Gray}
    %BTB 			 &33.73   	&22.85	&32.92  &34.86  &35.13	 &34.19  &33.01    &35.13    &32.64   &33.17   \\  \hline
		%
		%$T$        &3         &8      &9      &9      &3 	     &9      &9        &3        &8       &3        \\ %
  \end{tabular} 
	}
\end{table}
\vspace{-20pt}

% Retina $448 \times 832$, Chicken $160 \times 160$, Blueberry $256 \times 256$  

%\begin{figure*}[t]
%
        %\includegraphics[width=\linewidth]{fig3_alg3.eps}
				%\\
        %\includegraphics[width=\linewidth]{fig3_alg4.eps}
				%\\
        %\includegraphics[width=\linewidth]{fig3_ac.eps}
%
%\caption{Visualization of chicken muscle speckle suppression cross-sectional OCT image.
%The first and the second row show a sequence of
%images $\mathbf{x}_t,t=1,2,..5,$ of algorithm 3 and algorithm 4, respectively. 
%Right  - last column - shows mean image averaged over the rest of the images in the same row. 
%The third row show angular compounding images of $M=25,45,65,85,105,125$ tilted stage images \cite{Keahey:2023}}. 
%\label{fig3}
%\end{figure*}

\paragraph{Run Time}
Our proposed approach avoids the need to train and to implement algorithms on massive parallel hardware such as GPUs. Running time is relatively very low: $\sim$130ms for image of size $1024\times 1024$ on i7-1085H CPU using Matlab.
In comparison, as stated by the authors of \cite{Torem:2023}: The bottleneck of their proposed algorithm is iteratively running their denoising DNN for $T=1000$ iterations, where time complexity increases with image size and linearly with $T$. For the task of denoising Poissonian image intensities Torem $\&$ Ronen (2023) state their denoising runtime was $\sim$1
minute, per image on RTX 3090 GPU. The number of iterations indicated for our method is of 3 orders of magnitude less than comparable iterative algorithms  \cite{Romano:2017,Cohen:2021,Kawar:2021,Milanfar:2023}.
A clear advantage of our proposed method is, therefore, significantly lower computation complexity.

\section{Conclusions}\label{sec6}

We introduced an iterative denoising algorithm for noise of unknown level, in the absence of a known degradation model.
We have demonstrated the applicability of the proposed framework in the presence of coherent and incoherent noise settings.
In addition, we offered an implementation suitable specifically for speckle interference. 
Our algorithm is simple, easy to implement, does not require paired input-output examples for training, and exhibits significantly low computational complexity comparing with state-of-the-art denoisers. 
Future work may extend the proposed technique to image restoration tasks, as well as space-variant models that may take into consideration scattering and attenuation in the relevant applications. 
Extensions to other applications, such as, ultrasound imaging and polarization-sensitive OCT, may also be of scientific value. 
Future work can also explore unfolding our algorithms, thus training a denoiser for each iteration. 
A possible extension can allow a different (learned) denoiser for each iteration.

% Using one or two orders of magnitude less iterations.

%In a sense, deep neural nets (DNNs) are trained to seek these atoms, specifically CNNs, which are trained to find a set of filters to perform a classification or a regression task at hand. However, it is not yet currently clear how exactly are the net's weights (filters) picked. Does the net pick the optimal features? Do they have any practical or physical meaning? Or are they simply the features that are just a ``good enough" fit to produce the best answer? 
%In this paper we address some of these questions, by addressing a specific practical problem, simplifying its solution and analyzing the neural net by breaking it to its fundamental components, in order to perhaps have a better grasp as to how a neural net's weights are chosen and how does these ``choices" affect a trained net's ability to detect certain feature or be blinded to others.

%\section*{Acknowledgment}
%The authors thank the associate editor and the anonymous reviewers for their constructive comments and useful
%suggestions.

%\section*{References}
%

\appendix

\setcounter{section}{0}

\section{}\label{appA}

\textbf{Proofs} \\
Theorem \ref{Theorem1}. \\
The proof follows the outline of the Banach fixed-point theorem.
Denote $d(\mathbf{x}_m,\mathbf{x}_{m-1})=\|\mathbf{x}_m-\mathbf{x}_{m-1}\|, \ m\in \mathbb{N}$.
%First, we show that 
%\begin{equation*}
%d(f(\mathbf{x}_m),f(\mathbf{x}_{m-1})) < q d(\mathbf{x}_m,\mathbf{x}_{m-1}) \ \forall m\in\mathbb{N}, \ q \in [0,1), \ \rho=q/3.
%\end{equation*}
The ideal denoiser obeys,
\begin{equation}\label{A.1}
\|\mathbf{x}_{m}-\mathbf{x}_{m-1}\| \leq q^{m-1}(1+q) \|\mathbf{w}\|.
\end{equation}
This follows by induction and the assumption that $\| \mathbf{w}_{m+1} \| \leq q \|\mathbf{w}_{m}\|$, where $\mathbf{x}_m=\mathbf{x}^*+\mathbf{w}_m$, and $q\in[0,1)$. 
%Also it holds that,
%\begin{flalign}
%\nonumber
%q\|\mathbf{x}_{m-1}-\mathbf{x}_m\| & =
%q\|\mathbf{x}^*+\mathbf{w}_{m-1}-\mathbf{x}^*-\mathbf{w}_m\|=  q\|\mathbf{w}_{m-1}-\mathbf{w}_m\| 
%\\
%&
%\geq q\|\mathbf{w}_{m-1}\|-q\|\mathbf{w}_m\| \geq q\|\mathbf{w}_{m-1}\|-\|\mathbf{w}_m\|.
%\end{flalign}
%Therefore, $d(f(\mathbf{x}_m),f(\mathbf{x}_{m-1})) < q d(\mathbf{x}_m,\mathbf{x}_{m-1})$ holds when
%$2\|\mathbf{w}_m\|<q\|\mathbf{w}_{m-1}\|-\|\mathbf{w}_m\|$.
%Thus, it is required that $3\|\mathbf{w}_m\|<q\|\mathbf{w}_{m-1}\|$. By assumption $\|\mathbf{w}_m\|<\rho\|\mathbf{w}_{m-1}\|$. Therefore $\rho=q/3$. 
%By induction, $d(f(\mathbf{x}_{m+1}),f(\mathbf{x}_{m})) < q^m d(\mathbf{x}_m,\mathbf{x}_{m-1})$.
It is not sufficient for each prediction to become arbitrarily close to the preceding one for the algorithm to converge to a fixed point solution. Hence, we now show that $\{\mathbf{x}_{m}\}_{m \in \mathbb{N}}$ is a Cauchy sequence. Let $k,m\in\mathbb{N}$ such that $k>m$. 
\begin{flalign*}
\nonumber
d(\mathbf{x}_k,\mathbf{x}_m) 
&\leq d(\mathbf{x}_k,\mathbf{x}_{k-1})+d(\mathbf{x}_{k-1},\mathbf{x}_{k-2})+...+d(\mathbf{x}_{m+1},\mathbf{x}_m) \\
\nonumber
&\leq
\big(q^{k-1}+q^{k-2}+...+q^{m}\big)(1+q) \|\mathbf{w}\| \\
& \leq q^{m}(1+q)  \|\mathbf{w}\| \sum_{l=0}^{k-m-1}q^l
\leq q^{m}(1+q)  \|\mathbf{w}\| \sum_{l=0}^{\infty} q^l 
= \frac{q^{m}(1+q)}{1-q} \|\mathbf{w}\|.
\end{flalign*}
Since $q\in[0,1)$, we can find $N$ large enough such that for some $\varepsilon>0$, $q^N<\frac{\varepsilon(1-q)}{(1+q)\|w\|}$. Therefore, for $k,m>N$ we have
\begin{flalign}
\nonumber
d(\mathbf{x}_k,\mathbf{x}_m) \leq \frac{q^{m}}{1-q}(1+q)\|\mathbf{w}\|<\frac{\varepsilon(1-q)}{(1+q)\|\mathbf{w}\|} \frac{(1+q)\|\mathbf{w}\|}{1-q}=\varepsilon.
\end{flalign}
Therefore, the sequence $\{\mathbf{x}_m\}_{m\in\mathbb{N}}$ is a Cauchy sequence. And since $\|\mathbf{w}_m\|\leq q^m \|\mathbf{w}\|$, $\mathbf{x}_m \rightarrow \mathbf{x}^*$ such that $f(\mathbf{x}^*)=\mathbf{x}^*$ is a fixed-point of $f$.
%$f$ cannot have more than one fixed-point since a pair of fixed points $\mathbf{x}^*_1,\mathbf{x}^*_2$ would contradict the contraction $d(f(\mathbf{x}^*_1),f(\mathbf{x}^*_2))=d(\mathbf{x}^*_1,\mathbf{x}^*_2) < q d(\mathbf{x}_m,\mathbf{x}_{m-1})$.
\qed

Note that it is possible to prove that $f(\cdot)$ converges to a fixed point under a more strict assumption that $f(\cdot)$ forms a contraction mapping \cite{Williamson:1973}, yet this assumption is not necessarily true for most denoisers, and may be challenging to verify.

Theorem \ref{Theorem2}. \\
The proof follows a similar outline. 
For the simple iteration method, we have
\begin{flalign*}
\mathbf{x}_{m+1} & = (1-\mu_m)\mathbf{x}_m + \mu_m f(\mathbf{x}_m) \\
& =  
(1-\mu_m)(\mathbf{x}^*+\mathbf{w}^\mathrm{x}_m) + \mu_m (\mathbf{x}^*+\mathbf{w}^\mathrm{f}_m) \\
& =
\mathbf{x}^*+(1-\mu_m)\mathbf{w}^\mathrm{x}_m + \mu_m\mathbf{w}^\mathrm{f}_m.
\end{flalign*}
Hence, $\mathbf{w}^\mathrm{x}_{m+1}=(1-\mu_m)\mathbf{w}^\mathrm{x}_m + \mu_m \mathbf{w}^\mathrm{f}_m$.
Therefore,
\begin{flalign*}
\| \mathbf{w}^\mathrm{x}_{m+1} \| & = \|(1-\mu_m)\mathbf{w}^\mathrm{x}_m + \mu_m \mathbf{w}^\mathrm{f}_m\| \\
& 
\leq
(1-\mu_m)\|\mathbf{w}^\mathrm{x}_m\|+ \mu_m \|\mathbf{w}^\mathrm{f}_m\| \\
& 
\leq
(1-\mu_m+\mu_m q)\|\mathbf{w}^\mathrm{x}_m\|,
\end{flalign*}
since we assumed our denoiser obeys $\|\mathbf{w}^\mathrm{f}_m\|\leq q \|\mathbf{w}^\mathrm{x}_m\| \ , q \in[0,1)$.
Denote $\tilde{q}=(1-\mu_m+\mu_m q) \in [0,1)$. Since $\| \mathbf{w}^\mathrm{x}_{m+1} \| \leq \tilde{q} \|\mathbf{w}^\mathrm{x}_m \|$, the rest of the proof follows the exact proof outline for Theorem~\ref{Theorem1} above starting from (\ref{A.1}), where $\tilde{q}$ replaces $q$. 
\qed

\section{}\label{appB}
\textbf{Receptive Field Normalization} 

Most SC iterative solvers are slowed down by the use of one global threshold (bias) that is repetitively employed in every iteration to detect each local feature shift along the signal, or a predetermined constant local threshold \cite{Daubechies:2004,Beck:2009}. Applying a global threshold at each iteration, results in annihilation of weak expressions when the threshold is too high, while stronger expressions  cast a ``shadow'' over low-energy regions in the signal, which in turn, can be interpreted as false-positive support locations. On the other hand, if the threshold is very small, many iterations are required to compensate for false detections of early iterations. These issues are aggravated in the presence of noise, and in real-time applications due to model perturbations. The proposed remedy in \cite{Pereg:2021} was therefore to re-scale each data point by a locally focused data energy measure, before applying a threshold. In other words, each receptive field of the data is scaled with respect to the local energy. This way even when the data is inherently unbalanced, we can still use a common bias for all receptive fields, without requiring many iterations to detect the features support. 
Pereg et al. \cite{Pereg:2022} also suggested to incorporate 2D-RFN in an encoder-decoder RNN-based learning system, primarily for supervised learning of inverse problems. The 2D-RFN is applied prior to the RNN. Then, the latent space support normalized weights are multiplied with the non-normalized projection of the learned dictionary on the original input signal to regain the local energy. The RFN
signal's distribution is confined to a smaller set of values. Thus, RFN decreases the signal's entropy
and the size of the typical set associated with the input distribution \cite{Pereg:2023A}. It was also shown that as the noise level is increased RFN-RNN version suppresses noise better than the RNN system.

\textit{Definition 1 (2D Receptive Field Normalization Kernel)}: A kernel $h[k,l], \ k,l\in\mathbb{Z}$, can be referred to as a receptive field normalization kernel if 
\vspace{-0.5em}
\begin{enumerate}
\item The kernel is positive: $h[k,l] \geq 0 \quad \forall k,l$.
\item The kernel is symmetric: $h[k,l]=h[-k,-l] \quad \forall k,l$.
\item The kernel's global maximum is at its center: $h[0,0] \geq h[k,l] \quad \forall k,l \neq 0$.
\item The kernel's energy is finite: $ \sum_{k,l} {h[k,l]} < \infty$.
\end{enumerate}
\textit{Definition 2 (2D Receptive Field Normalization)}: We define the local weighted energy centered around $Y[k,l]$, a sample of a 2D observed signal $\mathbf{Y}\in\mathbb{R}^{M \times N}$,
\begin{equation}\label{B.1}
\sigma_{\mathrm{y}}[k,l] \triangleq \Bigg( \sum_{k',l'=-\frac{L_\mathrm{h}-1}{2}}^{\frac{L_\mathrm{h}-1}{2}} { h[k',l'] y^2[k-k',l-l'] } \Bigg) ^ \frac{1}{2},
\end{equation} 
where $h[k,l]$ is a RFN window function of size $L_\mathrm{h} \times L_\mathrm{h}$, $L_\mathrm{h} << \min(M,N)$ is an odd number of samples. For our application we used a truncated Gaussian-shaped window, but one can use any other window function depending on the application, such as: a rectangular window, Epanechnikov window, etc. %The choice of the normalization window and its length affects the thresholding parameters. If $h[k,l]$ is a rectangular window, then $\sigma_{\mathrm{y}}[k,l]$ is simply the Frobenius norm of a data patch centered around the $[k,l]$th location. Otherwise, if the chosen RFN window is attenuating, then the energy is focused in the center of the receptive field, and possible events at the margins are repressed.
RFN is employed by dividing each data point by the local weighted energy. 
We compute local weighted energy as defined in (\ref{B.1}). Namely,
\begin{equation}\label{B.2}
\sigma_{\mathrm{y}}[k,l]=\sqrt{h[k,l]*Y^2[k,l]},
\end{equation}
where $h[k,l]$ is a receptive field normalization window, and $*$ denotes the convolution operation.
Similarly to the 1D case, we normalize the signal by dividing each data point by the corresponding receptive field energy.
In order to avoid amplification of low energy regions, we use a clipped version of $\sigma_{\mathrm{y}}[k]$. Namely,
\begin{equation}\label{B.3}
\tilde{\sigma}_{\mathrm{y}}[k,l]=
\begin{cases}
\sigma_{\mathrm{y}}[k,l] &   \sigma_{\mathrm{y}}[k,l]  \geq \tau \\
1 &  \sigma_{\mathrm{y}}[k,l]  < \tau
\end{cases},
\end{equation}
where $\tau>0$ is a predetermined threshold. Empirically, for our application $0.15 \leq \tau \leq 0.4$ works well. 
The receptive-filed normalized image image $\tilde{\mathbf{Y}}$ is therefore,
\begin{equation}\label{B.4}
\tilde{\mathbf{Y}}[k,l]=\frac{\mathbf{Y}[k,l]}{\tilde{\sigma}_{\mathrm{y}}[k,l]} .
\end{equation}

\section{}\label{appC}
Proof of Theorem \ref{Theorem4}. \\

\begin{enumerate} 
\item When $\mathbf{v}=c\mathbf{1}$, and $h[k]=1/L_\mathrm{h}$,
\begin{equation*}
\sigma_{\mathrm{v}}[k] = \Bigg( \sum_{n=-\frac{L_\mathrm{h}-1}{2}}^{\frac{L_\mathrm{h}-1}{2}} { h[n] v^2[k-n] } \Bigg) ^ \frac{1}{2} = c.
\end{equation*} 
Assuming $c>\tau$,
\begin{equation*}
\tilde{v}[k]=v[k]/\tilde{\sigma}_\mathrm{v}[k]=1.
\end{equation*}
Therefore,
\begin{equation*}
g(\mathbf{v}) = (\tilde{\mathbf{v}}-1)\odot \mathbf{v} = \mathbf{0}.
\end{equation*}

\item Recall that,
\begin{equation*}
\sigma^2_{\mathrm{v}}[k] = 
\frac{1}{L_\mathrm{h}} \sum_{n=-\frac{L_\mathrm{h}-1}{2}}^{\frac{L_\mathrm{h}-1}{2}} {v^2[k-n]}.
\end{equation*} 
Therefore, for we have
\begin{flalign*}
E \sigma^2_{\mathrm{v}}[k] & = 
E \bigg\{ \frac{1}{L_\mathrm{h}} \sum_{n=-\frac{L_\mathrm{h}-1}{2}}^{\frac{L_\mathrm{h}-1}{2}} \bigg[ v^2[k-n] - m_\mathrm{v}^2  \bigg]+ m_\mathrm{v}^2 \bigg\}\\
& = m_\mathrm{v}^2 + \frac{1}{L_\mathrm{h}} \sum_{n=-\frac{L_\mathrm{h}-1}{2}}^{\frac{L_\mathrm{h}-1}{2}} 
E \big\{ v^2[k-n]-m_\mathrm{v}^2 \big\} = s^2_\mathrm{v} + m_\mathrm{v}^2.
\end{flalign*} 

For $m_{\mathrm{v}}=0$, assuming $s^2_\mathrm{v}>\tau$,
\begin{equation*}
E \tilde{v}^2[k] = E v^2[k]/s^2_\mathrm{v} \approx 1.
\end{equation*}
\end{enumerate}
\qed

\bibliography{dpereg_2023}

\end{document}